\documentclass[sigplan,10pt]{acmart}
\makeatletter
\def\@ACM@checkaffil{
    \if@ACM@instpresent\else
    \ClassWarningNoLine{\@classname}{No institution present for an affiliation}%
    \fi
    \if@ACM@citypresent\else
    \ClassWarningNoLine{\@classname}{No city present for an affiliation}%
    \fi
    \if@ACM@countrypresent\else
        \ClassWarningNoLine{\@classname}{No country present for an affiliation}%
    \fi
}
\makeatother

\acmConference[XXX'25]{XXX}{January 1--January 1, 2025}{XXX, XXX}
\acmYear{2026}
\acmISBN{} 
\acmDOI{} 
\startPage{1}

\setcopyright{none}

\usepackage{booktabs}   
\usepackage{subcaption} 
\usepackage{pifont}
\usepackage{enumitem}

\usepackage{color}
\usepackage{makecell}
\usepackage{multirow}
\usepackage{algorithm}
\usepackage{algpseudocode}

\newcommand{\ourmethod}{\textsc{EaaS}}

\begin{document}

\title[Expert-as-a-Service]{Expert-as-a-Service: Towards Efficient, Scalable, and Robust Large-scale MoE Serving}         


\author{Ziming Liu}
\affiliation{National University of Singapore}
\affiliation{Shanghai Qiji Zhifeng Co., Ltd.}
\email{liuziming@comp.nus.edu.sg}
\authornote{Work done during Ziming Liu's internship at Shanghai Qiji Zhifeng Co., Ltd.}

\author{Boyu Tian}
\affiliation{Shanghai Qiji Zhifeng Co., Ltd.}
\email{tianboyu@qijizhifeng.com}

\author{Guoteng Wang}
\affiliation{Shanghai Qiji Zhifeng Co., Ltd.}
\email{wangguoteng@qijizhifeng.com}

\author{Zhen Jiang}
\affiliation{Shanghai Qiji Zhifeng Co., Ltd.}
\email{jiangzhen@qijizhifeng.com}

\author{Peng Sun}
\affiliation{Shanghai Qiji Zhifeng Co., Ltd.}
\email{sunpeng@qijizhifeng.com}

\author{Zhenhua Han}
\affiliation{Shanghai Qiji Zhifeng Co., Ltd.}
\email{hanzhenhua@qijizhifeng.com}

\author{Tian Tang}
\affiliation{Shanghai Qiji Zhifeng Co., Ltd.}
\email{tangtian@qijizhifeng.com}

\author{Xiaohe Hu}
\affiliation{Infrawaves}
\email{huxiaohe@infrawaves.com}

\author{Yanmin Jia}
\affiliation{Infrawaves}
\email{jiayanmin@infrawaves.com}

\author{Yan Zhang}
\affiliation{Infrawaves}
\email{zhangyan@infrawaves.com}

\author{He Liu}
\affiliation{Infrawaves}
\email{liuhe@infrawaves.com}

\author{Mingjun Zhang}
\affiliation{Infrawaves}
\email{Mingjun.Zhang02@gmail.com}

\author{Yiqi Zhang}
\affiliation{National University of Singapore}
\email{yiqi.zhang@u.nus.edu}

\author{Qiaoling Chen}
\affiliation{Nanyang Technology University}
\email{qiaoling.chen@ntu.edu.sg}

\author{Shenggan Cheng}
\affiliation{National University of Singapore}
\email{shenggan@comp.nus.edu.sg}

\author{Mingyu Gao}
\affiliation{Tsinghua University}
\email{gaomy@tsinghua.edu.cn}

\author{Yang You}
\affiliation{National University of Singapore}
\email{youy@comp.nus.edu.sg}
\authornote{Corresponding authors.}

\author{Siyuan Feng}
\affiliation{Shanghai Innovation Institute}
\affiliation{Shanghai Qiji Zhifeng Co., Ltd.}
\email{syfeng@sii.edu.cn}
\authornotemark[2]

\fancyhead{}  
\renewcommand\footnotetextcopyrightpermission[1]{} 


\begin{abstract}
Mixture-of-Experts (MoE) models challenge serving infrastructures with dynamic, sparse expert utilization, causing instability on conventional systems designed for dense architectures. We propose \ourmethod{}, a novel serving system to enable efficient, scalable, and robust MoE deployment. Our system disaggregates MoE modules into independent, stateless services. This design enables fine-grained resource scaling and provides inherent fault tolerance by decoupling compute units. The architecture is powered by a high-performance, CPU-free peer-to-peer communication library that ensures minimal overhead and high throughput. Experiments confirm \ourmethod{}'s scalability and efficiency, achieving performance comparable to monolithic systems while providing robust fault tolerance and strong scalability. \ourmethod{} incurs less than a 2\% throughput reduction under simulated hardware failures that would otherwise halt monolithic architectures. It further saves up to 37.5\% of computing resources through dynamic fine-grained adaptation to serving traffic, demonstrating strong resilience for large-scale MoE deployment in production.
\end{abstract}

\maketitle

\section{Introduction}
\label{sec:introduction}

The rapid advancement of Large Language Models (LLMs) has revolutionized natural language processing, enabling unprecedented capabilities but simultaneously introducing significant computational challenges for deployment. Serving these increasingly large and complex models efficiently is paramount for their practical application. Mixture-of-Experts (MoE) architectures \cite{switch-transformers, qwen3, mixtral, llama4, deepseek-v3, deepseek-v2} have emerged as a promising direction, mitigating computational costs by leveraging sparsity. Unlike dense models where all parameters participate in every computation, MoE models activate only a subset of specialized sub-models (``experts'') per input token, potentially reducing computational load significantly while maintaining high performance. Furthermore, recent fine-grained-expert MoE models, such as the DeepSeek series~\cite{deepseek-v2, deepseek-v3, deepseek-r1}, push sparsity further by employing numerous smaller experts, enabling scaling to near trillions of parameters without a proportional increase in inference cost.

Despite the architectural advantages, efficiently serving MoE models presents unique challenges distinct from those of dense LLMs. While several frameworks (e.g., vLLM~\cite{vllm}, SGLang~\cite{sglang}, and TensorRT-LLM~\cite{trt-llm}) excel at serving dense models, they often struggle with the dynamic, sparse nature of MoE computation. First, existing systems can exhibit poor elasticity, requiring large, monolithic deployment units (e.g., DeepSeek-V3/R1 requires 320 GPUs as a basic scaling unit for high efficiency~\cite{deepseek-v3,deepseek-r1}), which suffer from coarse scaling granularities. 
Second, mainstream approaches rely heavily on large-scale collective communication primitives (like All-to-All for expert parallelism) established within static process groups at initialization. This tight coupling makes the system brittle: a single device failure can necessitate restarting the entire group. 
Third, the inherent dynamism of MoE leads to workload imbalance. The distribution of activated experts can vary drastically depending on the input data and downstream task, yet existing systems often employ static sharding strategies that cannot adapt, resulting in inefficient resource utilization where some expert-hosting devices are overloaded while others remain idle.
While recent attempts have explored~\cite{step-3, megascale-infer} disaggregating the attention and MoE layers, they often fail to resolve these fundamental issues, retaining static group structures and tightly coupled control logic that limits flexibility and robustness.

To address these limitations, we propose \ourmethod{}, a scalable, robust, and efficient serving system designed for MoE LLMs. The core principle of \ourmethod{} is to re-architect the system around the concept of disaggregating experts into independent services. We decouple the MoE layers from the rest of the model, such as the attention layers, treating the pool of experts as dynamically accessible and independent services. This fundamental shift from monolithic deployment to a service-oriented architecture offers several key advantages:

\textbf{Fine-grained Elasticity:} This service-oriented approach eliminates the rigid structure of large serving units. \ourmethod{} allows expert capacity to be scaled almost linearly and independently from the attention computation components. For instance, serving a model like DeepSeek-V3 \cite{deepseek-v3} can begin with a practical base unit (e.g., 16x 80GB GPUs) and scale incrementally, potentially one GPU at a time, to precisely match demand. Furthermore, \ourmethod{} removes the need for static communication group establishment between GPUs, enabling more precise scaling according to the current workload.

\textbf{Improved Robustness:} By replacing collective communication with peer-to-peer (P2P) interactions between attention clients and expert servers, \ourmethod{} removes the dependency on fragile, static communication groups for expert parallelism. If an expert server fails, clients can transparently failover to a replica or an alternative instance after a timeout with minimal disruption. Recovery is also simplified; since experts operate as services, a new server only needs to register its availability to be integrated into the system, avoiding the costly rebuild of a global communication group.

\textbf{Dynamic Load Balancing:} The independence of each expert server enables real-time load balancing. If specific experts are frequently requested by more tokens, they can be duplicated to balance the computation. Furthermore, these new expert instances can be added to the serving pool without interrupting the overall service, allowing the system to adjust to shifting workloads and prevent performance bottlenecks.

\begin{figure}[t]
    \centering
    \includegraphics[width=0.98\linewidth]{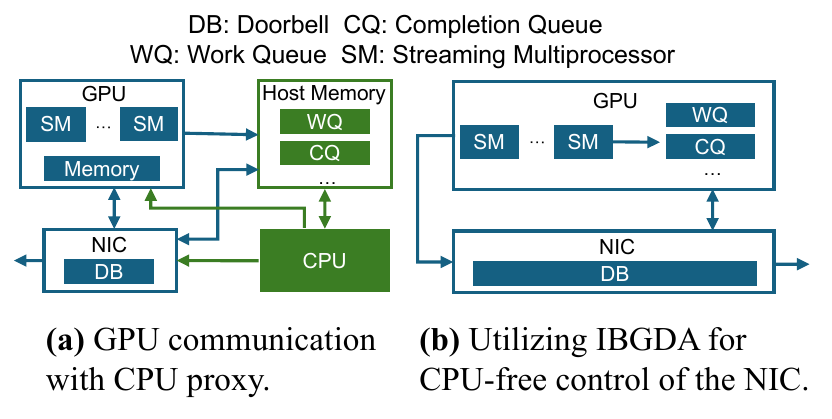}
    \caption{\ourmethod{} leverages InfiniBand GPUDirect Async (IBGDA) to achieve low communication latency and minimize kernel launch overhead through full CUDA graph capture, enabled by CPU-free control.}
    \label{fig:ibgda}
\end{figure}

Additionally, efficiently implementing this disaggregated model requires a flexible, high-performance communication substrate for dynamic, asymmetric P2P interactions. Existing libraries often fall short: NCCL~\cite{nccl} lacks suitable single-sided P2P operations, while RDMA-based libraries like NVSHEMEM~\cite{nvshmem} impose restrictive symmetric buffer requirements. While recent works such as StepMesh~\cite{step-3} and Megascale-Infer~\cite{megascale-infer} leverage CPU-controlled communication, introducing extra overhead and preventing end-to-end CUDA graph capture.
To overcome these limitations, we developed a novel asymmetric asynchronous P2P communication library using low-level IBGDA (InfiniBand GPUDirect Async) primitives. Our library offers flexible buffer management, native support for efficient single-sided operations, minimal overhead through CPU-free and group-free network operation, perfectly aligning with the requirements of  \ourmethod{}.

In summary, our contributions are:
\begin{enumerate}
    \item We propose \ourmethod{}, a novel serving system architecture for fine-grained MoE models based on disaggregating experts into independent services. This design enhances elasticity, robustness, and load balancing compared to monolithic approaches.  \label{contrib:system}
    \item We design and implement a high-performance, asymmetric, asynchronous CPU-free P2P communication library based on IBGDA, tailored for dynamic client-server interactions in GPU clusters, overcoming limitations of existing libraries for this use case. \label{contrib:comm_lib}
    \item Through extensive experiments, we show that \ourmethod{} achieves performance comparable to state-of-the-art monolithic systems while providing superior fault tolerance and scalability. \ourmethod{} shows less than 2\% decoding throughput drop during hardware failures that would halt conventional architectures and allows for incremental, cost-effective scaling not possible with rigid, group-based designs, saving up to 37.5\% computing resources.
    \label{contrib:evaluation}
\end{enumerate}
\section{Background and Motivation}
This section provides the necessary background on Mixture-of-Experts models and current serving systems. We then analyze the fundamental limitations of existing monolithic designs, which motivate our new approach.

\subsection{Mixture-of-Experts: Scaling with Sparsity}

Mixture-of-Experts (MoE) has become a dominant architecture for scaling Large Language Models efficiently. The core idea is to replace dense feed-forward network (FFN) layers with a larger set of smaller, specialized "expert" sub-networks. For each input token, a lightweight gating network dynamically selects a small subset of these experts to perform the computation. This conditional execution allows models to grow to trillions of parameters in total capacity without a proportional increase in the computational cost for any single inference pass.

\begin{figure}[t]
    \centering
    \begin{subfigure}[b]{0.48\linewidth}
        \centering
        \includegraphics[width=\linewidth]{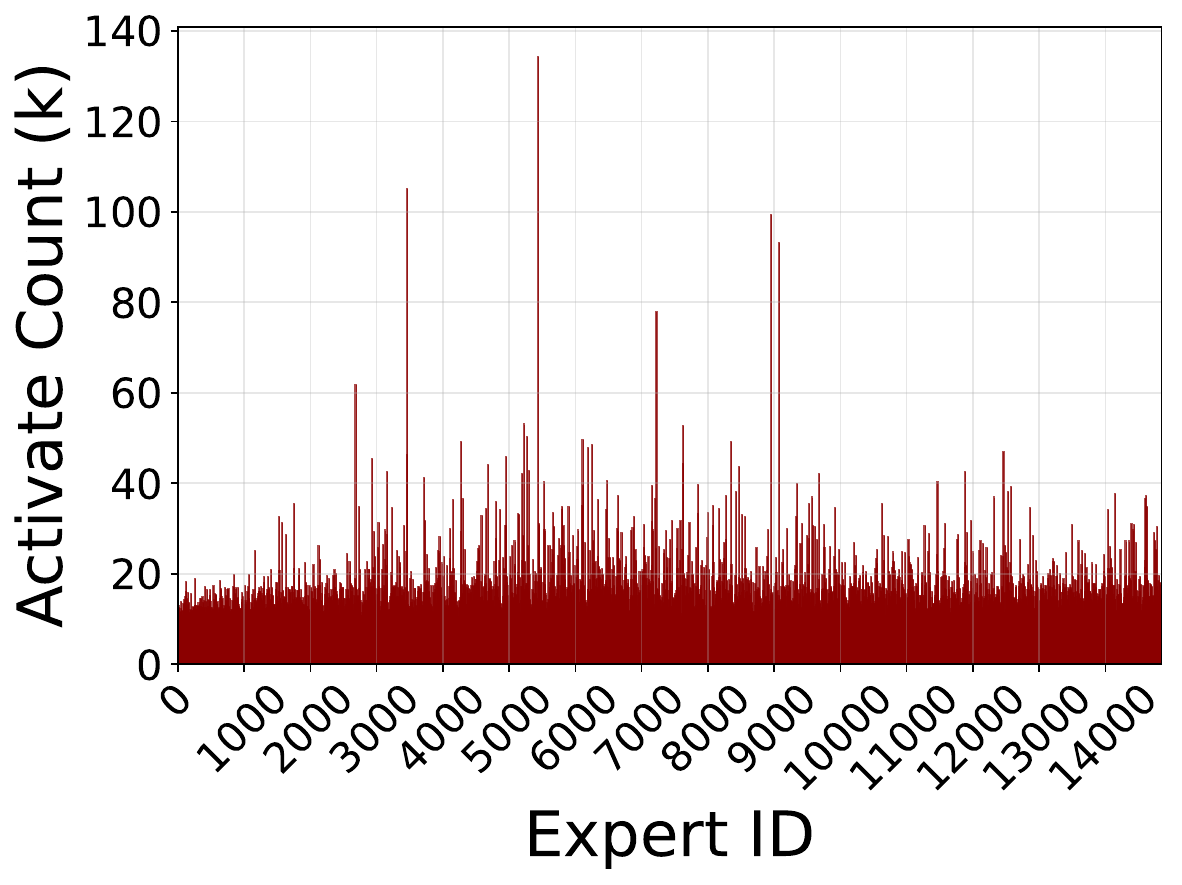}
        \caption{ShareGPT}
    \end{subfigure}
    \hfill
    \begin{subfigure}[b]{0.48\linewidth}
        \centering
        \includegraphics[width=\linewidth]{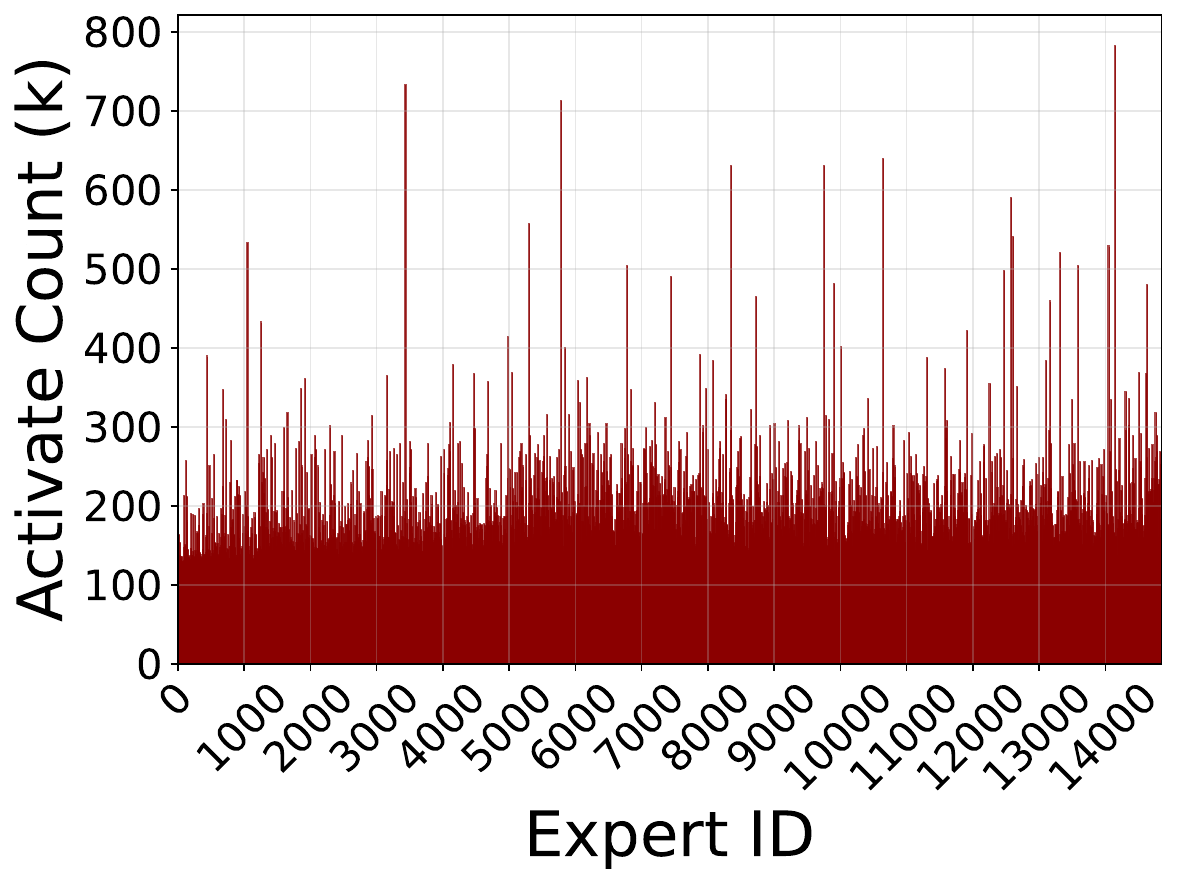}
        \caption{LongBench}
    \end{subfigure}
    
    \caption{The distribution of expert activations varies significantly across different downstream tasks, posing a dynamic load balancing challenge.}
    \label{fig:activation}
\end{figure}

Recent architectural trends have moved from models with a few large experts (e.g., Mixtral-8x7B~\cite{mixtral}) toward fine-grained MoE, featuring thousands of much smaller experts (e.g., DeepSeek-V3~\cite{deepseek-v3}, Kimi-K2~\cite{kimi-k2}, Qwen3~\cite{qwen3}). As shown in \autoref{tab:models}, this approach enables enormous model scale while keeping the number of activated parameters manageable. However, this fine-grained sparsity introduces two significant challenges for serving systems:

\begin{itemize}[leftmargin=*]
    \item \textbf{Low Computational Intensity}: During batched decoding, tokens are scattered across a vast number of experts. This means each expert processes only a small fraction of the tokens in a batch, leading to poor hardware utilization unless extremely large, often impractical, batch sizes are used.
    \item \textbf{Dynamic Workload Imbalance}: Token-level expert assignments are highly dynamic and input-dependent. As illustrated in \autoref{fig:activation}, the distribution of expert activations can vary dramatically across different tasks and datasets. This data-dependent skew creates a difficult load-balancing problem that static resource allocation schemes cannot solve.
\end{itemize}

\begin{table}[t]
    \centering
    \begin{tabular}{c|ccc}
    \Xhline{2\arrayrulewidth}
        \multirow{2}{*}{\centering \scalebox{0.8}{\textbf{Model}}}  & \scalebox{0.8}{\textbf{Total}} & \scalebox{0.8}{\textbf{Active}} & \scalebox{0.8}{\textbf{Min. Unit}}\\[-5pt]
        & \scalebox{0.8}{\textbf{Param. (B)}} & \scalebox{0.8}{\textbf{Param. (B)}} & \scalebox{0.8}{\textbf{(80GB GPUs)}}\\
    \hline
    Qwen3-235B-A22B  &  235 &  22 &  8 \\
    DeepSeek V3      &  671 &  37 & 16 \\
    Kimi-K2          & 1000 &  32 & 32 \\
    Llama 4 Behemoth & 2000 & 288 & 64 \\
    \Xhline{2\arrayrulewidth}
    \end{tabular}
    \vspace{3pt}
    \caption{While MoE models activate only a fraction of their total parameters, housing the entire model still demands substantial GPU resources, necessitating distributed serving.}
    \label{tab:models}
\end{table}

\subsection{The Monolithic MoE Serving Architecture}

To handle the immense size of modern MoE models (\autoref{tab:models}), distributed serving is a necessity. The predominant approach, adopted by systems like vLLM \cite{vllm}, SGLang \cite{sglang}, and TensorRT-LLM \cite{trt-llm}, can be characterized as a monolithic architecture. In this design, the entire model, including both attention and expert layers, is deployed as a single, tightly-coupled distributed application.

At the heart of this architecture lies \emph{Expert Parallelism (EP)}, a technique where experts are sharded across a group of GPUs. To route tokens from the attention layers to their designated experts, these systems rely heavily on large-scale \emph{collective communication} primitives, most notably \texttt{All-to-All}. During an MoE layer computation, each GPU sends the tokens destined for a specific expert to the GPU hosting it. This requires a collective data exchange where every GPU in the group communicates with every other GPU.

This reliance on collective communication forces the creation of a \emph{static process group} at initialization. All participating GPUs must be known beforehand, and the communication patterns are pre-determined. While effective for static workloads, this monolithic, group-based design creates a rigid structure that is ill-suited for the dynamic and demanding nature of production LLM serving. This rigidity gives rise to three fundamental limitations:

\begin{itemize}[leftmargin=*]
    \item \textbf{Coarse-Grained Elasticity.} The static process group is the indivisible unit of deployment and scaling. To add capacity, an entire new group of identical size and configuration must be provisioned. It's impossible to add just a few GPUs to handle a marginal increase in load. This leads to inefficient resource allocation, as systems must be over-provisioned for peak demand. The 320-GPU "basic scaling unit" recommended for DeepSeek-V3 \cite{deepseek-v3} is a clear example of this coarse granularity.

    \item \textbf{Brittleness and Low Fault Tolerance.} Collective communication is fragile. The failure of a single GPU or network link within the static group will cause the All-to-All operation to hang or fail, bringing the entire serving instance to a halt. Recovery requires restarting the entire group, a slow and disruptive process that reduces service availability.

    \item \textbf{Static and Inefficient Load Balancing.} The mapping of experts to GPUs is fixed when the service is launched. This static assignment cannot adapt to dynamic workload patterns, such as those shown in \autoref{fig:activation}. If certain experts become "hot" (frequently activated), the GPUs hosting them become performance bottlenecks, while GPUs with "cold" experts sit idle. The monolithic design lacks the flexibility to mitigate these imbalances by, for instance, dynamically replicating popular experts on underutilized hardware.
\end{itemize}

Recent works like MegaScale-Infer \cite{megascale-infer} and StepMesh \cite{step-3} have explored disaggregating the attention and MoE layers into separate tiers. However, they still rely on static, group-based collective communication between these tiers. They essentially create two coupled monolithic groups, thereby inheriting the same fundamental limitations of poor elasticity, brittleness, and static load management.

These challenges highlight a fundamental architectural mismatch between the static, rigid nature of monolithic serving systems and the dynamic, sparse computational patterns of fine-grained MoE models. This motivates our departure from the monolithic paradigm toward a more flexible, robust, and service-oriented design.

\begin{figure}[t]
    \centering
    \includegraphics[width=0.98\linewidth]{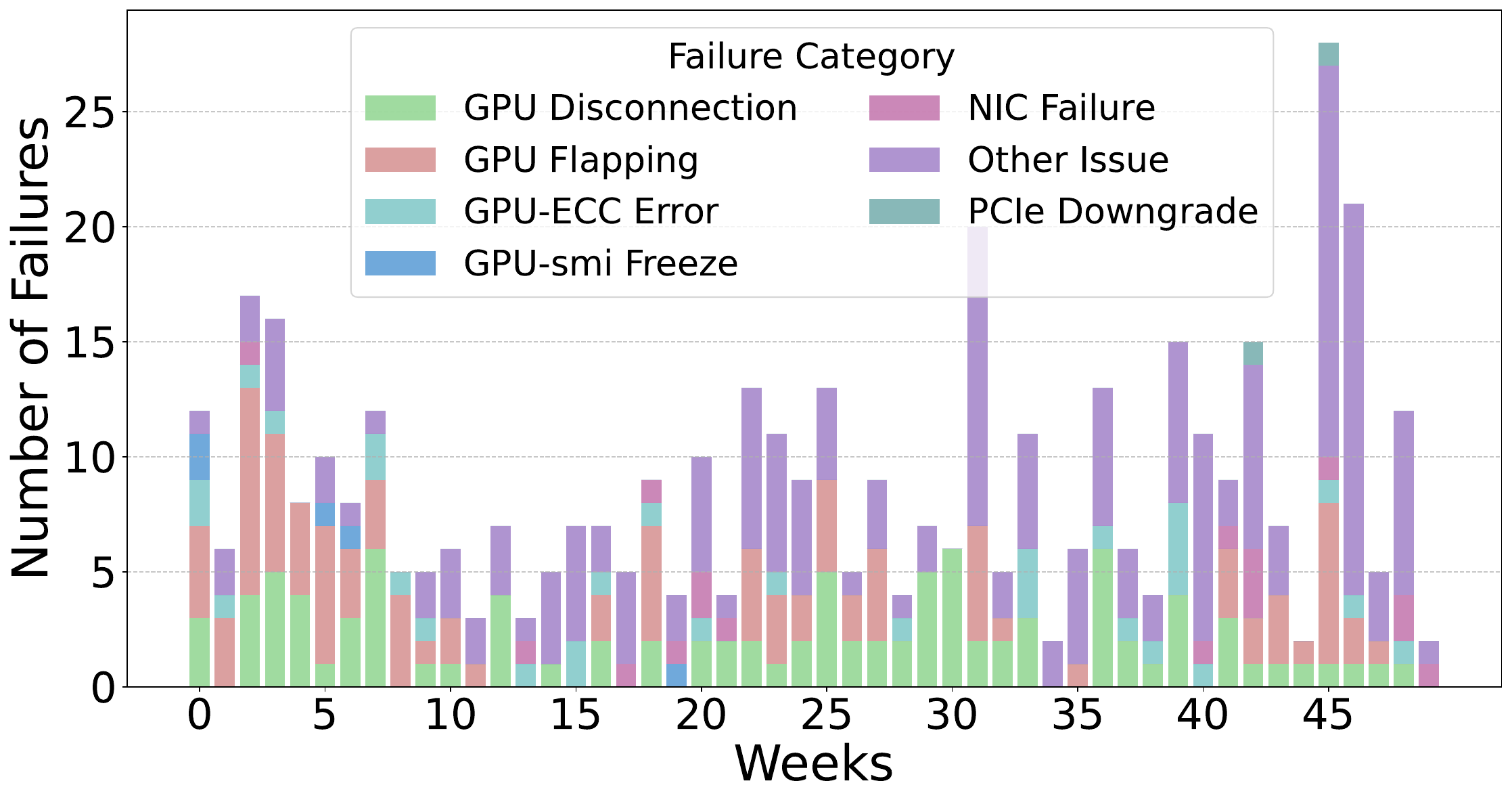}
    \caption{Weekly failure frequencies and categories observed in our computing cluster, consisting of 4096 NVIDIA Hopper GPUs with each node connected via 8×400 Gbps RoCE network. An average of 8.8 failures per week indicates that robust fault tolerance and rapid recovery are essential in such large-scale systems.
}
    \label{fig:failure}
\end{figure}

\subsection{Challenges of Efficiency and Scalability}

While the monolithic architecture is functional, its rigidity creates severe practical challenges when deployed in production environments, particularly for large-scale Model-as-a-Service (MaaS) providers. These challenges manifest as operational inefficiencies, poor reliability, and performance bottlenecks.

\paragraph{Inefficient Provisioning and High Operational Costs.}

The monolithic design, with its coarse-grained and indivisible scaling units, imposes a high "buy-in" cost and forces inefficient resource provisioning. MaaS providers must allocate resources for peak demand, meaning a large cluster of GPUs (e.g., the 320-GPU unit for DeepSeek-V3) remains committed even during periods of low traffic. This leads to significant resource stranding and drives up operational costs, as the provider pays for idle, tightly-coupled hardware. Furthermore, this inflexibility prevents providers from dynamically adjusting the ratio of attention-to-expert compute resources to match shifting workload characteristics, leading to a one-size-fits-all allocation that is rarely optimal.

\begin{figure*}[t]
    \centering
    
    \includegraphics[width=0.98\linewidth]{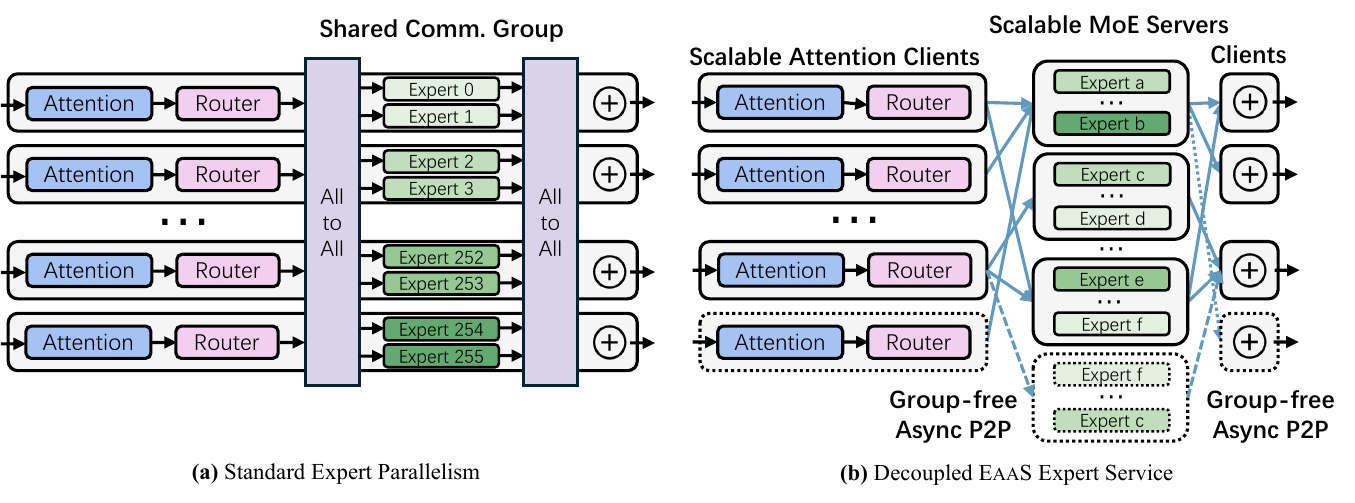}
    
    \caption{Overview of \ourmethod{} approach. Existing Expert Parallelism requires sharing communication groups with other collective parallelisms. \ourmethod{} disaggregates the MoE layers and other dense components, enabling flexible and scalable deployment of attention and MoE layer. Our approach extracts the experts in MoE layers as independent services. Once the attention clients finish the computation, the router will send the token to the corresponding expert server and gather the result back via async peer-to-peer connection, and then applies the weighted sum with router scores.}
    \label{fig:overview}
\end{figure*}

\paragraph{Poor Fault Tolerance and High Recovery Overhead.}
The tight coupling of resources within a static process group makes the entire serving instance extremely brittle. Collective communication primitives like \texttt{All-to-All} require every participant to be healthy; the failure of a single GPU or network link can cause the collective to hang indefinitely, triggering a cascading failure that brings down the entire service. As shown in \autoref{fig:failure}, hardware failures are a common occurrence in large-scale clusters. In a monolithic system, recovery from such an event is a costly, disruptive process, requiring the restart of the entire multi-hundred-GPU deployment. This leads to significant downtime, dropped user requests, and a severe impact on service-level objectives (SLOs).

\paragraph{Static Load Balancing and Performance Bottlenecks.}
The assignment of experts to specific GPUs is fixed at deployment time. This static allocation is fundamentally incapable of adapting to the dynamic, input-dependent nature of expert activation patterns (as illustrated in \autoref{fig:activation}). Consequently, when certain experts become "hot spots" due to the workload, the GPUs hosting them are overwhelmed, creating bottlenecks. Simultaneously, GPUs hosting "cold" or less-frequently used experts remain underutilized. The overall throughput is thus throttled by the single most overloaded device, not the aggregate capacity. This static imbalance prevents the system from achieving its potential performance and leads to inefficient use of expensive GPU resources.

\subsection{The Architectural Pivot: Exploiting the Stateless Nature of Expert}

The severe challenges of elasticity, robustness, and load balancing detailed above are not independent issues; they are symptoms of a single, fundamental architectural flaw in the monolithic design: the tight coupling of stateful and stateless computation. An LLM serving process has two distinct components:

\begin{itemize}[leftmargin=*]
    \item \textbf{Stateful Attention Layers:} These layers maintain the Key-Value (KV) cache, which stores the contextual state for each individual generation sequence. This component is inherently stateful and session-specific.
    \item \textbf{Stateless MoE Layers:} The expert networks within MoE layers perform a pure, idempotent function. They take a token's hidden state as input and produce an output, without retaining any memory or state from previous tokens in the same sequence.
\end{itemize}

The monolithic architecture shackles the stateless, flexible MoE layers to the stateful, rigid attention components. This forces the entire system to be managed as a single, stateful, indivisible unit, thereby inheriting the worst properties of both. This core insight—the statelessness of experts—presents a clear architectural pivot. By decoupling these components, we can design a system that treats experts as the independent, stateless services they truly are. This disaggregation directly unlocks solutions to the previously identified challenges:

\paragraph{Independent, Fine-Grained Scaling.} Decoupling allows the pool of expert servers to scale independently from the attention-processing frontends. Instead of provisioning rigid, monolithic 320-GPU blocks, a provider can add or remove expert-hosting GPUs one at a time to precisely match the computational demand for experts, eliminating the massive resource stranding and high operational costs associated with over-provisioning. The ratio of attention-to-expert compute can be dynamically tuned to suit the workload, rather than being fixed by a rigid hardware topology.

\paragraph{Robustness through Replication.} The stateless nature of experts makes them perfectly suited for replication. They become interchangeable, fungible resources. If an expert server fails, requests can be transparently rerouted to a replica without disrupting the stateful KV cache on the attention GPU. This replaces the brittle, all-or-nothing failure model of collective communication with a robust, gracefully degrading service model. Recovery is as simple as a new expert server registering its availability, eliminating the need for a costly, service-halting restart of the entire group.

\paragraph{Dynamic Load Balancing via Service Instantiation.} A service-oriented design breaks the static mapping of experts to specific GPUs. When the system detects that certain experts have become "hot spots" (as shown in \autoref{fig:activation}), it can dynamically instantiate new copies of those specific experts on underutilized GPUs in the cluster. These new instances can immediately begin serving requests, distributing the load and alleviating bottlenecks in real-time. This transforms load management from a static, pre-deployment guess into a dynamic, adaptive system optimization.

This architectural pivot from a monolithic, tightly coupled system to a disaggregated, service-oriented one forms the cornerstone of our approach. Based on these principles, we designed \ourmethod{}, a system for large-scale MoE model serving that is inherently elastic, robust, and efficient. We will detail its design in the following section.

\section{System Design}
\begin{figure*}[t]
    \centering
    \includegraphics[width=0.98\linewidth]{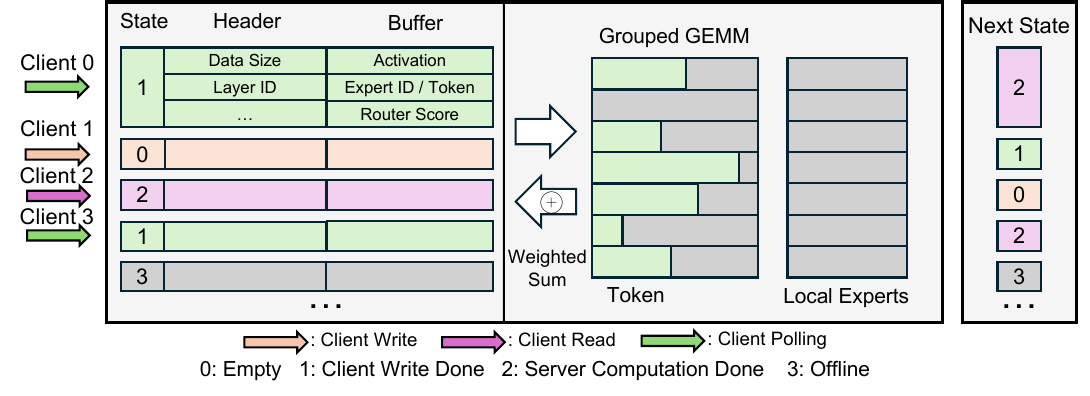}
    \caption{Dynamic batching for \ourmethod{} expert servers. The expert server continuously monitors and aggregates incoming requests from attention clients whose buffer state is "1" (indicating the client has finished writing and the data is ready for server-side computation). To ensure high GPU utilization, the server waits to aggregate a sufficient number of client requests before processing them as a batch. Tokens are then reorganized and processed using grouped GEMM operations. Upon completion of computation, the server writes the results back to the corresponding locations in the shared communication buffers and updates the buffer state to "2" (signaling that computation is finished and the output is ready for the client to read). Both read and write operations are performed as one-sided RDMA operations initiated by the client; the server does not initiate RDMA operations, allowing it to remain stateless and focused solely on computation.}
    \label{fig:server}
\end{figure*}

This section details the architecture of \ourmethod{}, a serving system designed for large-scale MoE models. It presents the system's disaggregated architecture, the shared communication buffer design, and the asynchronous expert server operation, which is critical for achieving scalability and robustness.

\subsection{Architectural Overview}

As depicted in \autoref{fig:overview}, the \ourmethod{} framework represents a paradigm shift for MoE model inference. Unlike conventional expert parallelism (\autoref{fig:overview} (a)), which uses a monolithic deployment that tightly couples attention and expert layers within a static group, \ourmethod{} employs a disaggregated, service-oriented architecture (\autoref{fig:overview} (b)). This design is rigid, vulnerable to single-point failures, and scales suboptimally.

The \ourmethod{} architecture decouples the stateless MoE layers from stateful components like the attention layers, creating two independently scalable entities: \textbf{1) Attention Clients:} These units handle dense computations like the attention mechanism. When an MoE layer is reached, the client's router dispatches activations to the appropriate expert servers via asynchronous peer-to-peer (P2P) communication. \textbf{2) Scalable Expert Servers:} Each server hosts a subset of experts and operates as an independent, stateless service focused solely on executing expert computations.

This architectural decoupling obviates the need for a shared, static communication group, which facilitates fine-grained elasticity and enhanced fault tolerance. Consequently, both clients and servers can be scaled independently to meet dynamic workloads, ensuring that an individual component failure does not cause a system-wide outage.

\subsection{Shared Communication Buffer Design}
The interaction between clients and expert servers is orchestrated through a shared communication buffer designed for efficient, asynchronous data exchange. As shown in \autoref{fig:server}, each client is allocated a dedicated buffer slot on the server, which is composed of a state flag, a header, and a data payload.

\textbf{Buffer State:} A single-byte flag denotes the buffer's status to mediate client-server interaction. The states are "0" (Empty) for client writing, "1" (Client Write Done) for server processing, 2 (Server Computation Done) for client reading, and "3" (Offline) to mark an inactive client.

\textbf{Header:} This section contains metadata for request processing, including the \texttt{layer\_id} and \texttt{batch\_size}. It can also include data size for memory management.

\textbf{Data Payload:} This section holds the core data for computation, including the token activation (hidden states), the \texttt{expert\_id} mapping for each token, and the \texttt{router\_score} for weighting outputs.

This structured buffer design creates a clear separation of responsibilities, where the client manages writing requests and reading results, while the server only reads requests and writes results. The state flag ensures proper synchronization without direct communication.

\subsection{Asynchronous Server Operation}

A fundamental principle of \ourmethod{} is the complete operational independence and asynchronicity of the expert servers. It is critical to note that the server does not initiate any synchronization or direct communication with the clients. Instead, it functions within a continuous, stateless loop, responding exclusively to the state of the shared communication buffers.

The server's operational workflow proceeds through a sequence of distinct steps. First, the server continuously and asynchronously polls the state flags of all client buffers allocated to it. Upon identifying all buffers in the "1" (Client Write Done) state, it aggregates these ready-to-process requests from multiple clients to form a dynamic batch. Subsequently, the tokens from this aggregated batch are reorganized and processed with high efficiency, utilizing optimized kernels such as grouped GEMM to achieve maximum hardware utilization. Following the completion of the computation, the server writes the results—specifically, the weighted sum of expert outputs—back into the data payload section of the corresponding client buffers. As a final step, the server updates the state flag of each processed buffer to "2" (Server Computation Done), thereby signaling to the respective clients that the results are available for retrieval.

This fully independent mode of operation is integral to the system's robustness and scalability. The stateless nature of the server and its independence from a static client group render it resilient to client failures. New servers can be dynamically integrated into the pool to manage increased loads, and clients can perform transparent failover to replica servers without service interruption. This design effectively transforms the expert modules into a scalable, robust, and on-demand computational service.

\begin{figure}[t]
    \centering
    \includegraphics[width=0.98\linewidth]{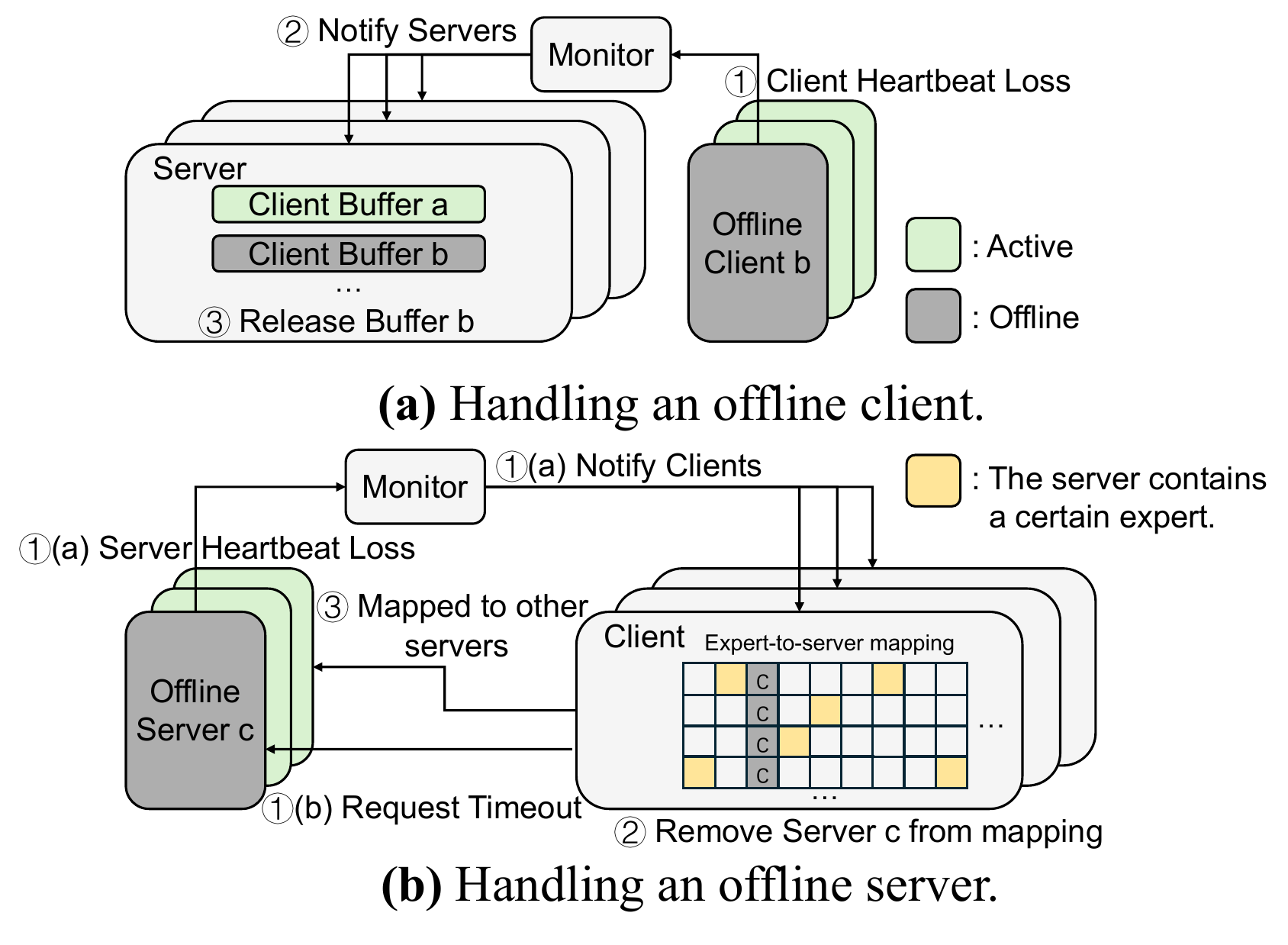}
    \caption{\ourmethod{} utilizes a monitor to listen to the heartbeat of all running workers and send notifications for servers to release buffer or for clients to modify the expert-to-server mapping mask when detecting an offline client/server. A client can detect an offline server either through the monitor (\ding{172}(a)) or a timeout request (\ding{172}(b)).}
    \label{fig:offline}
\end{figure}

\subsection{Fault Tolerance Mechanisms}
The disaggregated and asynchronous nature of the \ourmethod{} architecture, supported by a central monitor, provides inherent fault tolerance and ensures high service availability. As illustrated in \autoref{fig:offline}, the system is designed to handle failures of both clients and servers gracefully using a heartbeat-based health tracking mechanism.

\textbf{Client Failure:} In the event that an attention client goes offline, the monitor detects the loss of the client's heartbeat signal. It subsequently notifies all relevant expert servers of the failure. Upon receiving this notification, the servers release the communication buffer allocated to the offline client, preventing resource locks and ensuring that the failure remains localized. This event only affects the inference requests being processed by that specific client; all other clients and servers continue their operations without interruption, preserving overall system stability.

\textbf{Server Failure:} The system's resilience to server failure is achieved through the duplication of stateless expert services and is handled via two detection pathways. First, the central monitor can detect a server's heartbeat loss and proactively notify all clients. Second, a client can independently infer a server failure if a request's polling period exceeds a predefined timeout threshold. In either case, the client immediately updates its local expert-to-server mapping to remove the failed server from its list of available resources. It then automatically re-transmits the original request to an alternative server that hosts a replica of the required expert. This failover mechanism is designed to be transparent and have a negligible impact on the final inference outcome, ensuring continuous and robust service delivery.
\section{Optimizations}

\ourmethod{} is an integrated, end-to-end MoE serving system that combines SGLang~\cite{sglang}-based attention clients, an RDMA-based asymmetric asynchronous peer-to-peer (P2P) communication library based on IBGDA~\cite{IBGDA}, and a self-designed~\cite{openai-triton} MoE server. Custom Triton and CUDA kernels have been developed to reduce CPU involvement and to utilize CUDA Graph~\cite{cuda-graph} for decreasing kernel launch overhead. 

\begin{algorithm}[t]
\small
\caption{Iterate tokens with various length of batches.}
\begin{algorithmic}[1]
\Require $num\_tokens$: Array of token counts per batch
\Require $batch\_size$: Number of batches
\Require $compute\_func$: The function to process a single token

\State \textbf{Alloc shared memory:} $s\_num\_tokens[MAX\_BATCH]$
\State \textbf{Initialize:} $token\_id \gets blockIdx.x$
\State \textbf{Load }$s\_num\_tokens \gets num\_tokens$

\For{$batch\_id = 0$ \textbf{to} $batch\_size - 1$}
    \State $num\_token\_this\_batch \gets s\_num\_tokens[batch\_id]$
    \While{$token\_id < num\_token\_this\_batch$}
        \State \textbf{Call} $compute\_func(batch\_id, token\_id)$
        \State $token\_id \gets token\_id + gridDim.x$
    \EndWhile
    \State $token\_id \gets token\_id - num\_token\_this\_batch$
\EndFor

\end{algorithmic}
\label{algo:mask_kernel}
\end{algorithm}

\subsection{Kernel Optimization for Expert Server}

A key challenge in the \ourmethod{} architecture is managing the non-uniform workloads from imbalanced batches in large-scale MoE serving, which often leads to underutilized GPUs. We address this with specialized kernel optimizations (\autoref{algo:mask_kernel}) that use a static GPU grid. Each block independently iterates over valid tokens, skips empty batches, and uses shared memory to distribute tokens in a strided fashion. This design effectively balances workloads across GPU threads, allowing servers to adapt to imbalances without idle threads or synchronization overhead.

A second challenge arises from hosting many experts on each server. The standard \texttt{DeepGEMM} implementation creates performance bottlenecks by inefficiently iterating over all potential expert groups, most of which are inactive, and incurring a separate, low-throughput global memory read for each one. To resolve this, we introduce a \texttt{group-shrink} kernel that uses a GPU prefix scan to isolate active groups and move their metadata to the front so that we can early-stop. The \texttt{DeepGEMM} scheduler is then modified to iterate over this compacted tensor, which is loaded into shared memory once to eliminate inefficient, repeated memory reads.

These kernel-level optimizations, in concert with our dynamic batching and token reorganization strategies, enable expert servers to sustain high GPU utilization and throughput. By effectively managing these dynamic and sparse MoE workloads, our system overcomes the traditional bottlenecks of load imbalance and resource underutilization.

\subsection{Double-Batch-Overlap in Attention Client}
In \ourmethod{} architecture, a critical performance bottleneck is the communication latency inherent in dispatching tokens from attention clients to remote expert servers. This network round-trip time can force the client's GPU into an idle state while it awaits results, creating bubbles in the execution pipeline that directly degrade overall throughput. To counteract this inefficiency, we implement the Double-Batch-Overlap~\cite{deepseek-v3} strategy, a pipelining technique specifically designed to overlap this communication and remote computation latency.

This strategy ensures continuous GPU engagement by maintaining two active batches in the client's pipeline. The process unfolds as follows: while a previously dispatched batch (Batch A) is being processed by the remote expert servers, the client's GPU does not wait. Instead, it immediately commences the computationally intensive attention calculations for the subsequent batch (Batch B). This temporal overlap is the key mechanism; the attention computation for Batch B effectively hides the network and processing latency of Batch A. As a result, by the time the expert results for Batch A are returned to the client, Batch B is fully prepared for dispatch. This seamless handover creates a continuous, overlapping workflow that transforms potential idle periods into productive computation. By minimizing GPU stalls, this approach directly increases resource utilization, sustains a higher operational intensity for improved system throughput, and reduces the overall end-to-end latency for each request.

\subsection{CUDA Graph for End-to-End Acceleration}
High overhead from launching numerous small CUDA kernels can limit LLM inference performance. CUDA graphs address this by capturing the entire sequence of kernel launches into a single, reusable unit, which reduces CPU launch overhead.

Our use of a CPU-free communication library based on IBGDA is instrumental in maximizing the benefits of this technique. Because all network operations are initiated from the GPU without CPU intervention, we can capture the complete end-to-end workflow, from attention computation and data dispatch to result reception and final processing, within a single CUDA graph on the attention client. This holistic capture eliminates nearly all CPU-side launch overhead for the client's entire operational loop.

By enabling graph-based execution on both clients and servers, \ourmethod{} minimizes kernel launch overhead across the distributed system, enhancing overall efficiency and reducing latency.

\begin{figure}
    \centering
    \includegraphics[width=0.98\linewidth]{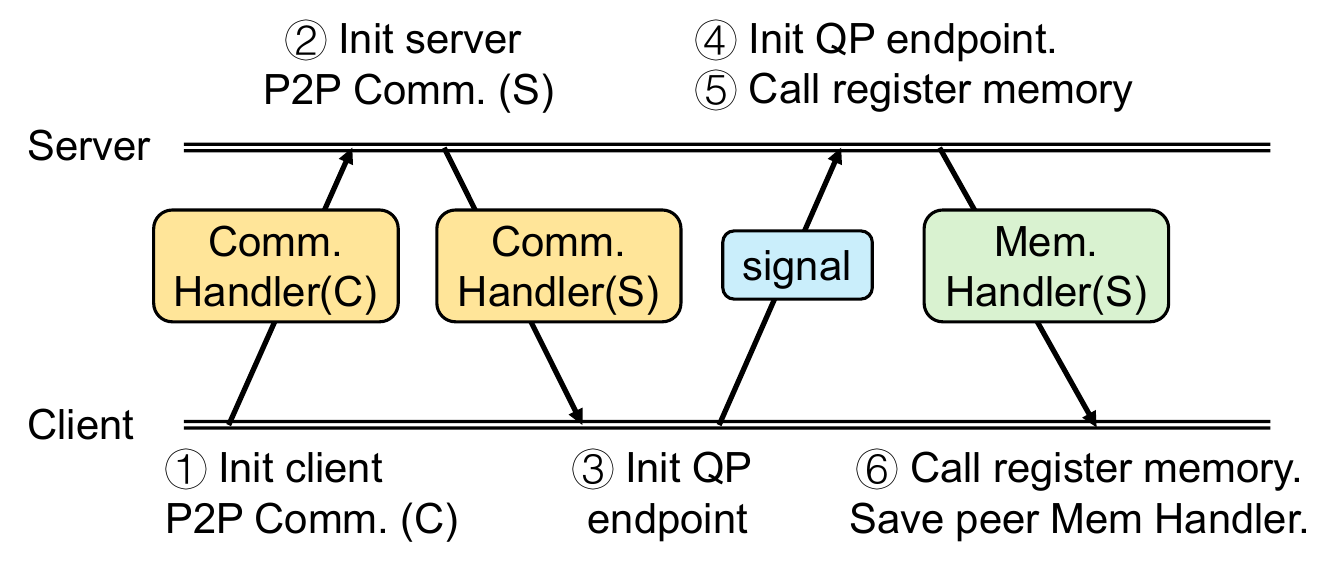}
    \caption{The process of establishing IBGDA connection}
    \label{fig:ibgda_init}
\end{figure}

\subsection{Flexible IBGDA Connection}
We utilize a client-server architecture to manage IBGDA connections, which ensures flexible deployment of \ourmethod{} services. The process for establishing a new IBGDA connection is illustrated in \autoref{fig:ibgda_init}. Firstly, the client initiates an HTTP request to the server \ding{172}, transmitting its IBGDA P2P communication handler (P2P Comm). Upon receiving this request, the server performs a mirroring operation \ding{173}. Meanwhile, the client begins to change its QP state and notifies the server \ding{174}\ding{175}. Finally, the client and server complete the registration of the GPU buffer and exchange their respective MR and buffer addresses \ding{176}\ding{177}, marking the peer's readiness state in their mappings.

To maintain a unified global view of the connection state, we introduce a central monitor. This monitor tracks the health of all clients and servers and broadcasts when a rank comes online or goes offline. It can be implemented using a distributed consistency component like ZooKeeper~\cite{ZooKeeper}. \autoref{fig:offline}  illustrates the system's fault-tolerance mechanisms for both client and server failures. When a server stops serving, the monitor's heartbeat mechanism detects the change, allowing its clients to update their mappings and switch to another available server. Conversely, a server will release the buffer of any client that has lost its heartbeat. This enables changing the number of instance without requiring a restart. Thanks to the monitor and client-server based IBGDA connections, \ourmethod{} only needs to rebuild connections on demand. This avoids the costly overhead of rebuilding the entire communication group during dynamic scaling or fault tolerance events.

\subsection{Dynamic Load Balancing for Experts}
\ourmethod{} is compatible with current mainstream MoE load balancing schemes, such as EPLB~\cite{eplb}, which balances the load among instances of expert parallelism by reordering and adding redundant experts. Furthermore, \ourmethod{} can use existing load-balancing strategies to provide a wider array of balancing methods. Specifically, three categories are possible. First, unlike expert parallelism, \ourmethod{} does not restrict the number of experts on each expert server to the same. This provides an opportunity for load balancing by adjusting the number of experts on each server. Second, based on the flexible IBGDA connection mechanism discussed above, \ourmethod{} can dynamically balance the load by scaling up or down the number of service instances for 'hot' and 'cold' experts. Finally, due to the decoupled deployment architecture, it is also possible to balance the load by changing the compute/memory specifications of the instances on which the experts are hosted. In conclusion, on top of existing load-balancing strategies, \ourmethod{} provides a richer set of load-balancing methods for deploying MoE model inference services, particularly in scenarios like cloud services. We will also explore load-balancing algorithms specifically designed for \ourmethod{} in future works.

\section{Evaluation}
In this section, we evaluate the performance of \ourmethod{} in comparison with state-of-the-art solutions for large-scale MoE model serving. Our experiments are designed to address the following questions:

\begin{itemize}[noitemsep, topsep=0pt]
    
    \item What is the overall performance of \ourmethod{} in realistic serving scenarios? (\S\ref{sec:eval:e2e})
    \item How is \ourmethod{}'s scalability with various amounts of workloads? (\S\ref{sec:eval:scalability})
    \item Does \ourmethod{} maintain acceptable performance during the recovery period after GPU failures? (\S\ref{sec:eval:recovery})
    \item Is the CPU-free design necessary for MoE serving? How efficient is the P2P communication library proposed in this work, compared with existing solutions? (\S\ref{sec:eval:comm})
    \item How much do key design components and implemented techniques contribute to the system’s overall performance? (\S\ref{sec:eval:ablation})
\end{itemize}

\begin{figure}[t]
    \centering
    \begin{subfigure}{\linewidth}
        \centering
        \includegraphics[width=0.9\linewidth]{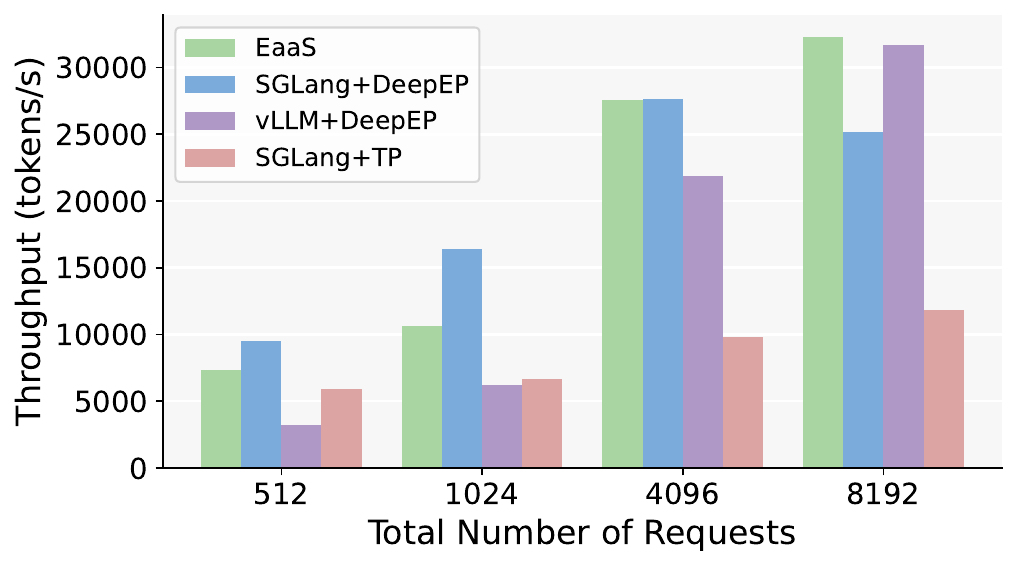}
        \caption{128 GPUs (64 prefilling and 64 decoding).}
    \end{subfigure}

    \vskip\baselineskip
    \begin{subfigure}{\linewidth}
        \centering
        \includegraphics[width=0.9\linewidth]{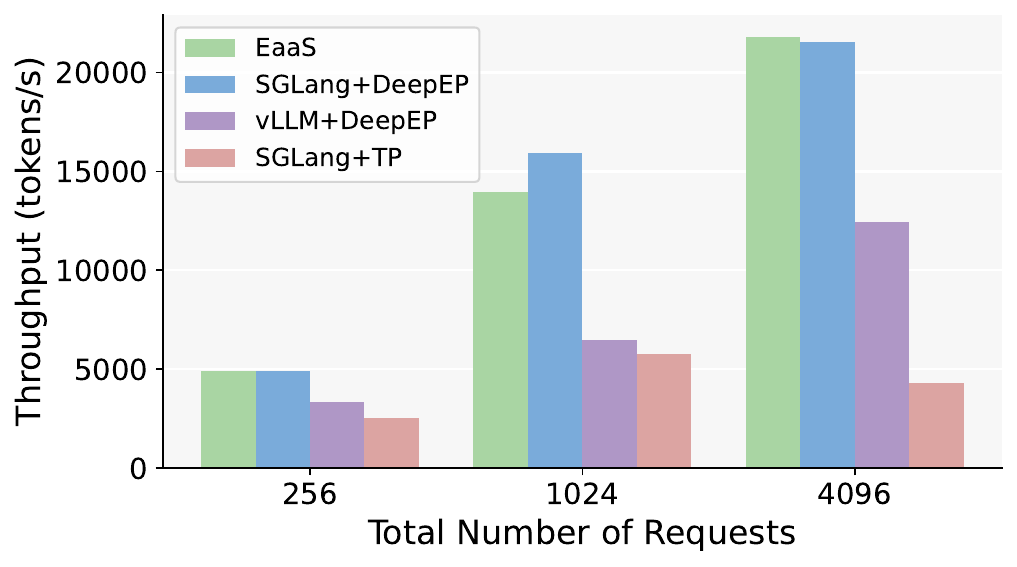}
        \caption{48 GPUs (16 prefilling and 32 decoding).}
    \end{subfigure}

    \caption{This figure compares the decoding throughput of \ourmethod{} against baseline systems across various request loads for two different GPU cluster sizes. The results demonstrate \ourmethod{'s} competitive end-to-end performance and scalability.}
    \label{fig:throughput}
\end{figure}

\begin{figure}[t]
    \centering
    \begin{subfigure}{\linewidth}
        \centering
        \includegraphics[width=0.9\linewidth]{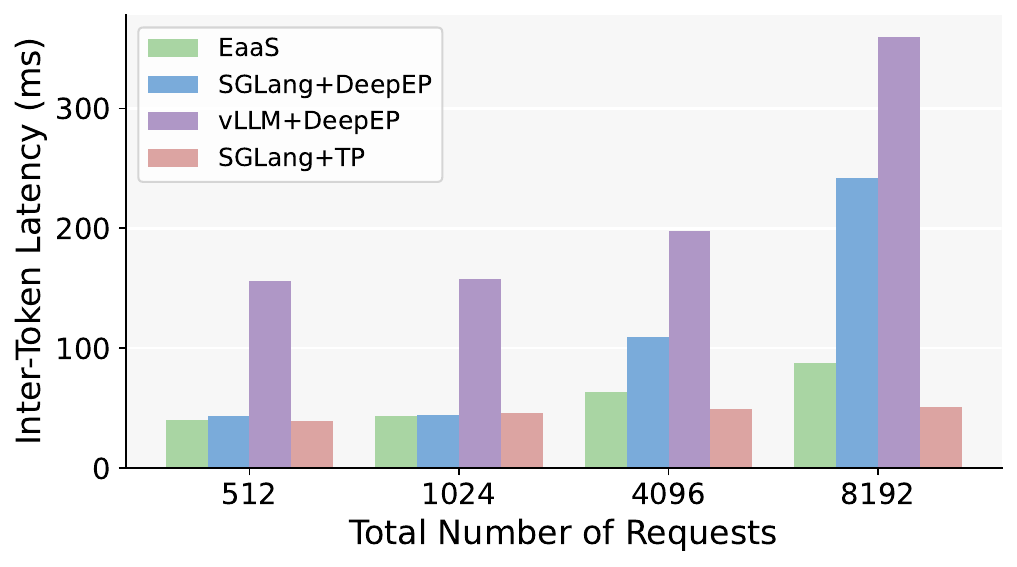}
        \caption{128 GPUs (64 prefilling and 64 decoding).}
    \end{subfigure}

    \vskip\baselineskip
    \begin{subfigure}{\linewidth}
        \centering
        \includegraphics[width=0.9\linewidth]{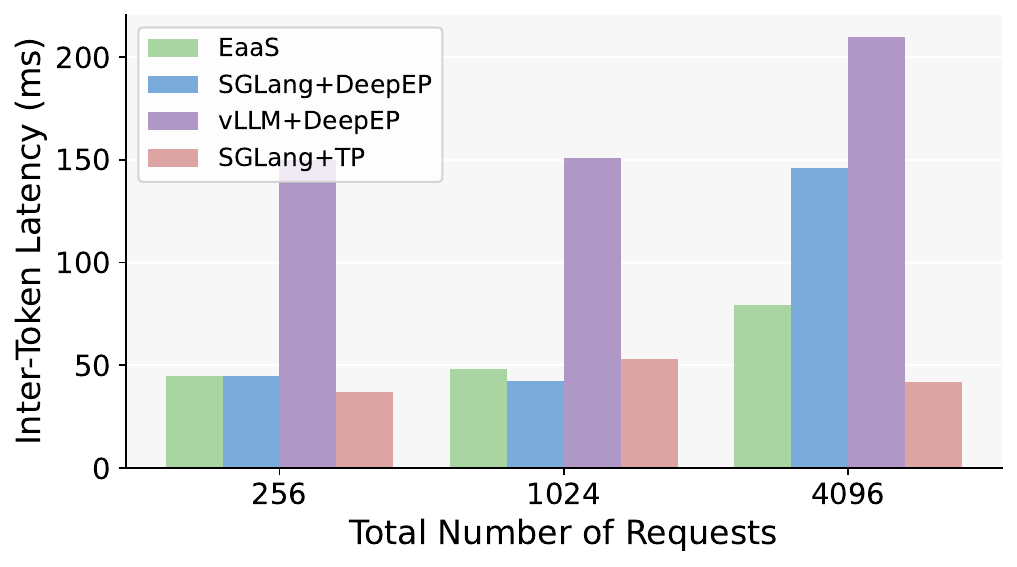}
        \caption{48 GPUs (16 prefilling and 32 decoding).}
    \end{subfigure}

    \caption{\ourmethod{} successfully balances high throughput with low inter-token latency, a key challenge in large-scale model serving. The system's architecture allows it to achieve strong processing speeds while maintaining acceptable responsiveness for users. Evaluations confirm that \ourmethod{} delivers competitive throughput and latency compared to state-of-the-art systems.}
    \label{fig:itl}
\end{figure}

\subsection{Evaluation Setup}
\textbf{Hardware and frameworks.} We use 16 computing nodes, each equipped with 8 Nvidia Hopper GPUs (80GB), connected via NVLink within the node and 8$\times$400Gbps RoCE links across nodes. For frameworks, we use PyTorch 2.7.1, Python 3.12, and CUDA 12.6.

\textbf{Model and workloads.} We evaluate \ourmethod{} on the DeepSeek-R1~\cite{deepseek-r1} 671B model under large-scale scenarios. For workloads, we adopt ShareGPT~\cite{sharegpt}. To mitigate long-tail effects, we cap the maximum response length at 768 tokens.

\textbf{Baselines.} We compare \ourmethod{} with the state-of-the-art inference engine SGLang~\cite{sglang} (commit ID: 51cdd81f) and vLLM~\cite{vllm} (commit ID: 9b01870). We also adopt relevant techniques from the SGLang codebase (e.g., KV caching) that are orthogonal to our contributions. Both SGLang and \ourmethod{} implement PD (Prefill-Decode) disaggregation with Mooncake~\cite{qin2025mooncakekvcachecentricdisaggregatedarchitecture}, while vLLM does not use disaggregation as this feature of its is currently experimental. SGLang supports two serving modes for DeepSeek models: Tensor Parallelism (TP) and Expert Parallelism (EP) via DeepEP~\cite{deepseek-v3}, a high-performance communication library. We include both modes to provide a comprehensive comparison across different communication patterns. Similarly, we use vLLM+DeepEP as another baseline. Since the same expert load balancing algorithm yields different expert distributions under \ourmethod{} and the baselines due to their distinct arrangements, we do not incorporate load balancing in the main evaluations. Instead, we include it in the ablation study to examine \ourmethod{}’s compatibility with these algorithms.

In addition, MegaScale-Infer~\cite{megascale-infer} and StepMesh~\cite{stepfun2025step3largeaffordablemodelsystem} propose asymmetric communication libraries based on GDRCopy~\cite{GDRCopy}, which rely on the CPU during communication. Since MegaScale-Infer's code is not publicly available, we compare our end-to-end communication overhead against StepMesh to highlight the necessity of our CPU-free design.

\textbf{Key metrics.} We measure two primary metrics: throughput and inter-token latency (ITL). For throughput, we focus on the decoding phase, as \ourmethod{} is applied to disaggregated decoding. We provision sufficient prefilling resources to ensure decoding throughput is not bottlenecked. ITL is measured to assess the ability of each method to deliver acceptable per-request latency.All metrics are measured using SGLang’s original serving benchmark scripts, and we report the end-to-end results across all submitted requests.

\subsection{End-to-End Evaluation}
\label{sec:eval:e2e}
We evaluate the overall performance of \ourmethod{} under two settings: (1) 16 nodes (128 GPUs) with 8 nodes for prefilling and 8 nodes for decoding, and (2) 6 nodes (48 GPUs) with 2 nodes for prefilling and 4 nodes for decoding, as shown in \autoref{fig:throughput} and \autoref{fig:itl}. 

For the decoding stage, SGLang+DeepEP (referred to as SGL-EP) employs full Expert Parallelism (EP sizes of 64 and 32, respectively). In contrast, \ourmethod{} adopts a client-server architecture with 32 clients–32 servers and 16 clients–16 servers. SGLang+TP (referred to as SGL-TP) is constrained to units of 16 GPUs due to the limitation of tensor parallelism. Consequently, the model must be duplicated 4 times and 2 times in the two settings when using TP. 

SGL-EP achieves competitive throughput and inter-token latency when the request volume is relatively small. However, under heavy traffic, we observe pronounced long-tail latencies, which we attribute to limitations in its request transportation and load balancing across the Data Parallel Attention~\cite{dpattn} ranks. In contrast, SGL-TP maintains low and stable ITL due to its small inference unit size (16 GPUs per unit). Yet, this small unit size also requires multiple replications of the model weights, reducing the available memory per GPU and limiting the maximum batch size. As shown in \autoref{fig:throughput}, this results in substantially lower throughput compared to other methods. 

The current PD disaggregation in vLLM is still experimental and relatively unstable. Thus, for vLLM, we allocate the same nodes for prefilling and decoding and, for fairness, evaluate decoding on 64 and 32 GPUs while excluding prefilling overhead. From the results, vLLM delivers high throughput under large-scale configurations but suffers from higher inter-token latency relative to other methods. 

Finally, \ourmethod{} benefits from its efficient communication library and balanced resource allocation between attention and expert computation. As a result, it achieves strong throughput while maintaining acceptable latency. Overall, \ourmethod{} demonstrates competitive end-to-end performance compared to state-of-the-art systems such as SGLang and vLLM, while additionally offering features like fault tolerance, which we will evaluate in the following sections.

\begin{figure}[t]
    \centering
    \includegraphics[width=0.9\linewidth]{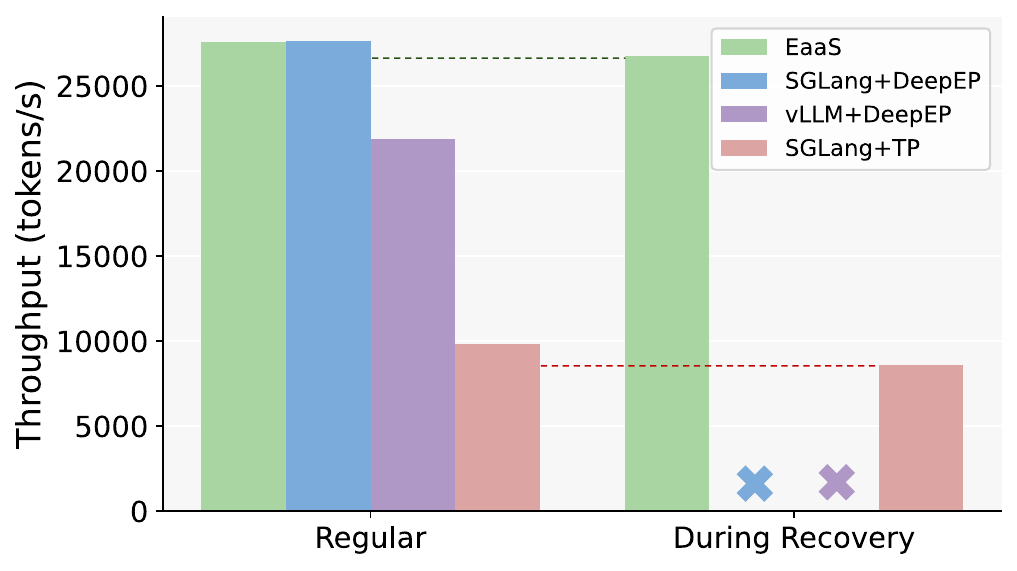}
    \caption{In our system fault tolerance experiment, \ourmethod{} proved resilient by maintaining high throughput without crashing, exhibiting only a slight performance decrease during recovery. DeepEP solutions, however, cannot continue operating under such failures. Because their monolithic architecture depends on a single communication group, the loss of one GPU forces the entire system to restart, causing a total service outage.}
    \label{fig:fault}
\end{figure}

\subsection{Scalability}
\label{sec:eval:scalability}
\begin{figure}[t]
    \centering
    \includegraphics[width=0.9\linewidth]{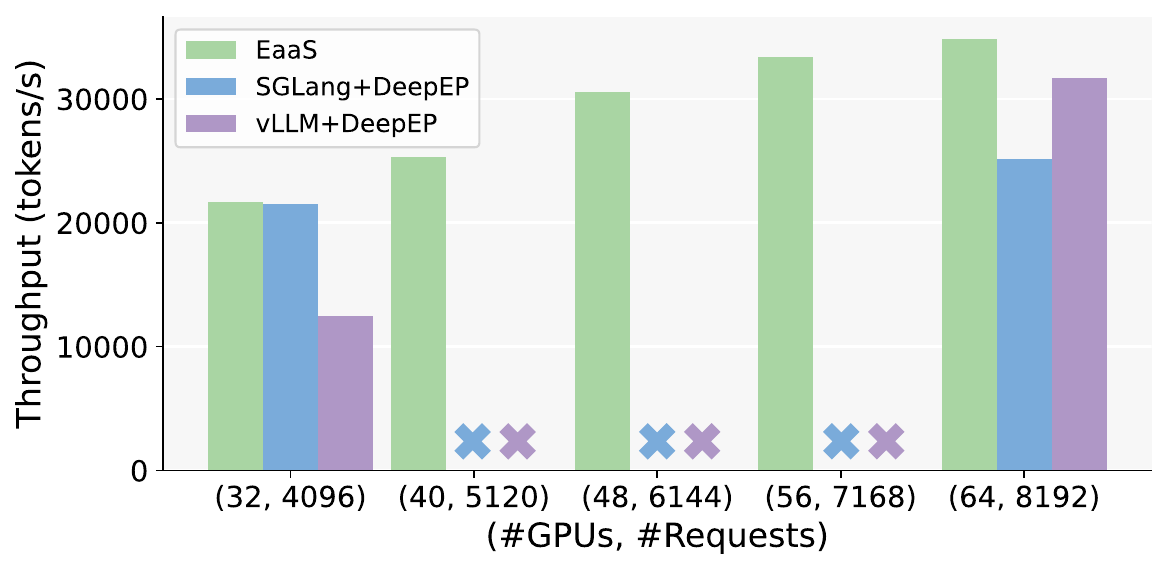}
    \caption{Weak scaling with fine granularity. SGLang and vLLM can only scale with GPU counts that evenly divide 7168, whereas \ourmethod{} supports arbitrary GPU counts, enabling more flexible and cost-efficient scaling.}
    \label{fig:scale}
\end{figure}

The scalability of \ourmethod{} was evaluated through a weak scaling experiment designed to demonstrate its capacity for dynamic, cost-effective resource allocation. As illustrated in Figure 11, the disaggregated architecture of \ourmethod{} exhibits a proportional, near-linear increase in throughput when scaling from 32 to 64 GPUs, indicating efficient resource utilization without significant overhead. This fine-grained scalability presents a considerable advantage over monolithic systems such as SGL-EP and vLLM-EP. These baseline systems are unable to operate on configurations with non-standard GPU counts (e.g., 40, 48, 56) due to their inherent architectural rigidity. Such inflexibility constrains operators to inefficient, coarse-grained scaling increments, resulting in resource stranding and elevated operational expenditures. Conversely, the service-oriented design of \ourmethod{} facilitates precise, demand-based provisioning by allowing the number of GPUs to be dynamically adjusted in response to workload fluctuations. For example, when the traffic decreases from 8192 to 5120, SGL-EP and vLLM-EP must still provision 64 decoding GPUs. In contrast, \ourmethod{} can scale down to 40 or 48 GPUs while maintaining the same level of decoding throughput as the baselines, thereby saving up to 37.5\% of computing resources.

\subsection{Fault Tolerance}
\label{sec:eval:recovery}
To evaluate fault-tolerance capabilities, we randomly disable ten GPUs, one at a time, and measure the average decoding throughput during recovery. The experimental setup follows the previous section, using 64 GPUs for decoding with a total of 4096 requests. As shown in \autoref{fig:fault}, both SGL-EP and vLLM-EP are unable to continue serving during recovery because all decoding GPUs belong to a single communication group; thus, the failure of one GPU necessitates restarting the entire group. In contrast, SGL-TP benefits from its smaller execution unit: only a group of 16 GPUs must be restarted upon a failure, while the remaining units remain operational. 

\ourmethod{} further improves resilience by pre-duplicating experts on servers, enabling expert requests originally routed to an offline server to be redirected to another server with a backup copy. Moreover, since attention clients in \ourmethod{} operate independently, client failures can also be efficiently handled by the routing mechanism of the PD disaggregation engine. Consequently, \ourmethod{} incurs less than 2\% drop in decoding throughput when a GPU failure occurs.

\subsection{Comparison with Other Asymmetric Communication Libraries}
\label{sec:eval:comm}
To evaluate communication library performance in realistic scenarios, we construct PyTorch tensors with the same shape used during the decoding phase of DeepSeek R1 models: (batch size, 1, 7168). In this setup, the client sends the tensor to the server, which immediately returns it to the client. We measure the end-to-end latency from the initiation of the client transmission to the completion of the response. Two scenarios are considered: a symmetric setting (2 clients and 2 servers) and an asymmetric setting (1 client and 3 servers). 

\begin{figure}[t]
    \centering
    \begin{subfigure}{\linewidth}
        \centering
        \includegraphics[width=0.9\linewidth]{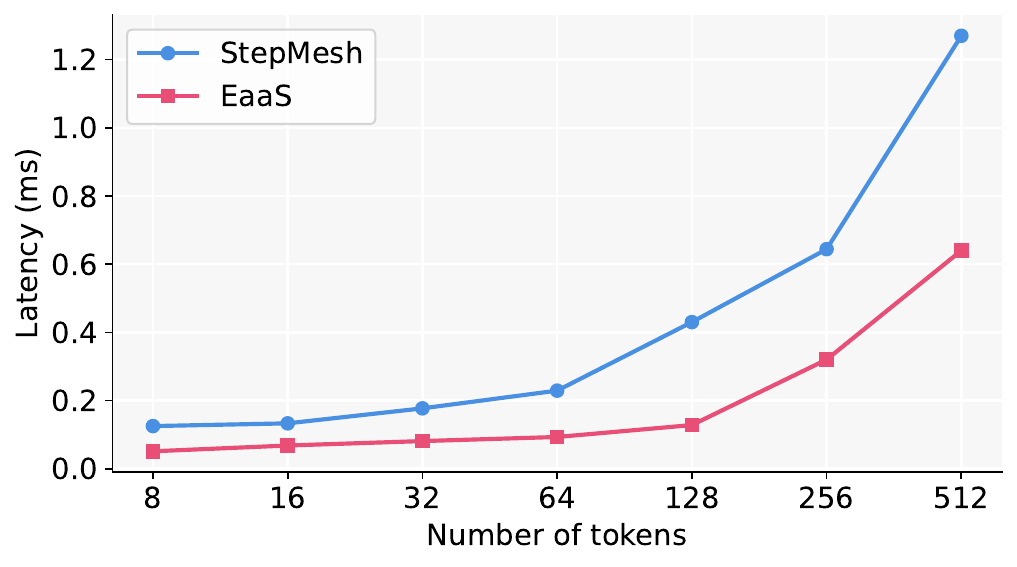}
        \caption{Latency under symmetric scenario.}
    \end{subfigure}

    \vskip\baselineskip
    \begin{subfigure}{\linewidth}
        \centering
        \includegraphics[width=0.9\linewidth]{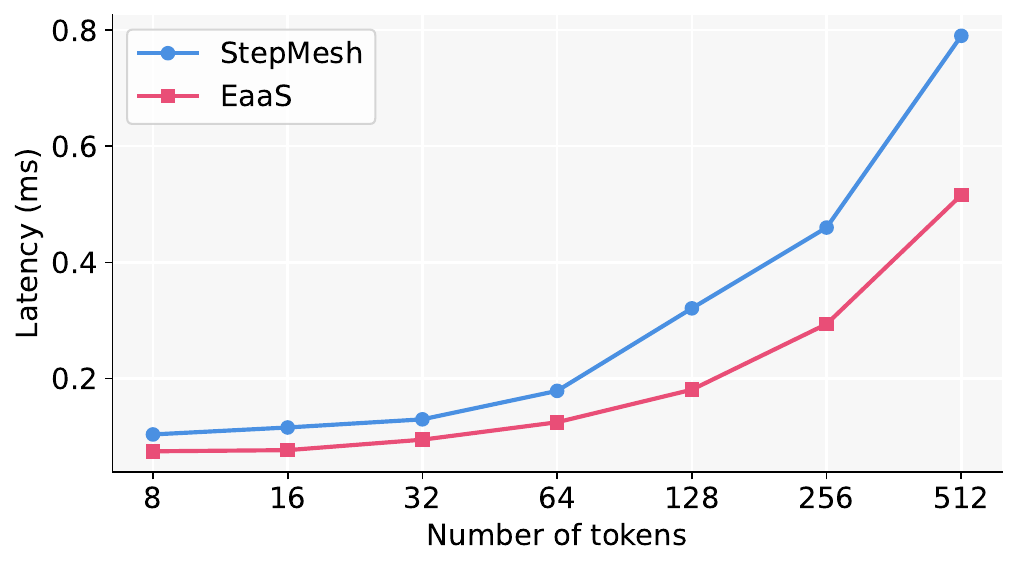}
        \caption{Latency under asymmetric scenario.}
    \end{subfigure}

    \caption{End-to-end overhead of round-trip communication, measured from sending a tensor via the communication API to receiving the returned result.}
    \label{fig:exp:comm_lib}
\end{figure}

As shown in \autoref{fig:exp:comm_lib}, \ourmethod{} consistently achieves lower latency than StepMesh, reducing overhead by 49.6\% and 34.7\% in the two scenarios when the batch size is 512. The performance gains of \ourmethod{} stem from two key factors. First, by leveraging IBGDA to bypass the CPU, we eliminate the overhead associated with accessing host memory queues and CPU-issued control orders. Second, as the process is entirely CPU-free, the launch overhead of communication kernels can be further reduced through CUDA graph capture. 

These results highlight the importance of CPU-free communication libraries, which existing systems such as StepMesh and MegaScale-Infer have yet to provide. More importantly, they demonstrate that reducing communication overhead at the library level directly translates into lower end-to-end latency for large-scale inference serving, thereby improving user-perceived responsiveness in real deployments.

\subsection{Ablation Study}
\label{sec:eval:ablation}

To better understand the contribution of each key component in \ourmethod{}, we conduct an ablation study by selectively disabling them and measuring the resulting throughput. Specifically, we examine three optimizations: (1) CUDA Graph capture, which reduces kernel launch overhead by eliminating CPU involvement; (2) kernel shrinking, which compacts group metadata to improve expert server efficiency; and (3) double batching, which overlaps attention computation with communication to mask latency.

\begin{figure}[t]
    \centering
    \includegraphics[width=0.9\linewidth]{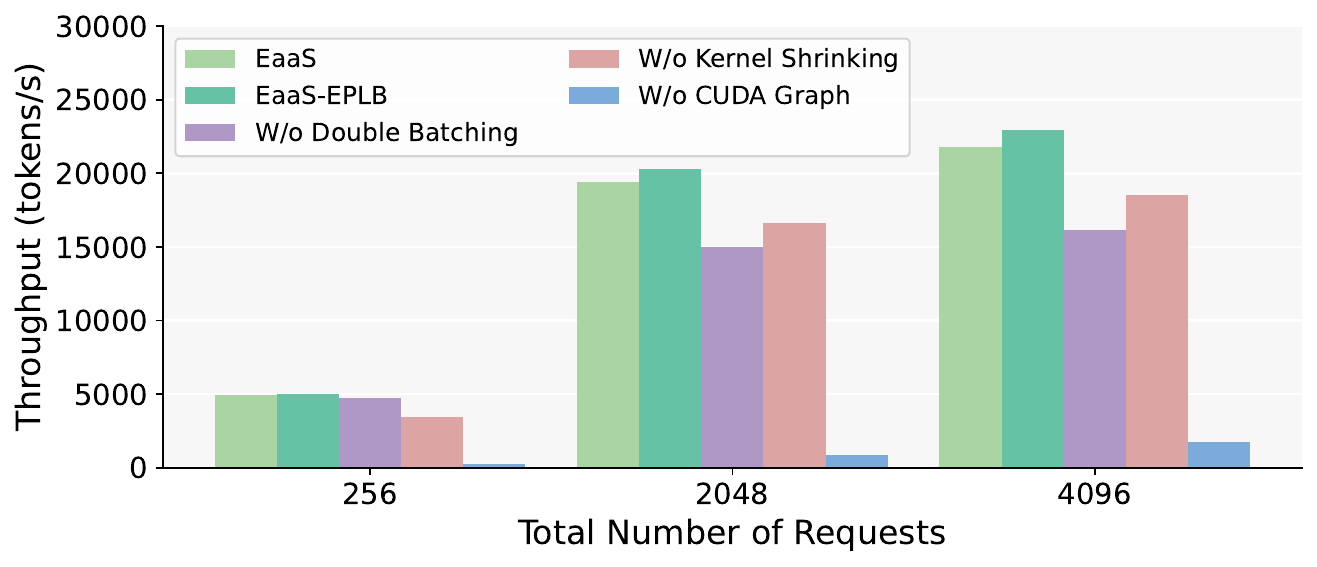}
    \caption{This ablation study demonstrates that \ourmethod{'s} key optimizations are all critical for performance, as disabling any of them causes a severe drop in system throughput.}
    \label{fig:ablation}
\end{figure}

As shown in \autoref{fig:ablation}, removing any of these optimizations results in a clear performance degradation. Using batch size $4096$ as an example, disabling CUDA Graph causes a dramatic \textbf{91.9\%} throughput drop due to repeated CPU-side kernel launch overhead. Without kernel shrinking, throughput decreases by \textbf{14.9\%} because of inefficient scheduling over sparse expert groups. Finally, removing the double-batching strategy leads to a \textbf{25.9\%} decline, as communication latency is no longer masked by overlapping attention computation. Together, these results confirm that all three optimizations are critical for achieving the overall efficiency of \ourmethod{}.

We also evaluated \ourmethod{} in combination with the expert load balancing algorithm proposed by DeepSeek~\cite{deepseek-v3}. The results show performance improvements of 2.4\%, 4.4\%, and 5.1\%, respectively. In future work, we plan to investigate expert load balancing strategies specifically tailored for \ourmethod{}.

\section{Related Works}

\paragraph{LLM Serving Frameworks} 
For efficient LLM serving, several frameworks have been proposed. vLLM \cite{vllm} introduces PagedAttention for efficient KV cache memory management. SGLang~\cite{sglang} achieves efficient serving with RadixAttention for prefix caching. MLC-LLM~\cite{mlc-llm} enables serving on diverse platforms through a machine learning compiler, while TensorRT-LLM~\cite{trt-llm} provides flexible LLM APIs with state-of-the-art optimizations and emphasizes ease of use. These frameworks are optimized for dense models and do not address the dynamic sparsity characteristic of MoE models. In this work, we leverage the inherent sparsity in MoE models to develop an efficient and robust serving system tailored specifically for them.

\paragraph{Parallelisms For LLMs}
To facilitate large-scale LLM inference and training, various forms of parallelism such as Tensor Parallelism (TP)~\cite{shoeybi2020megatronlmtrainingmultibillionparameter}, Pipeline Parallelism (PP)~\cite{gpipe, gems, li2021chimera, 10.1145/3581784.3607073, Lin2025WeiPipe}, and Data Parallelism (DP)~\cite{rajbhandari2020zeromemoryoptimizationstraining, zhao2023pytorchfsdpexperiencesscaling} have been used. In adapting the sparsity inherent in MoE models, Expert Parallelism (EP) was introduced to distribute experts across devices. In this work, we decouple EP from the dense portion of the model and from other parallelism dimensions, thereby enabling flexible scheduling and enhanced robustness. This decoupling permits group-free and resilient MoE serving.

Recent works such as Megascale-Infer~\cite{megascale-infer} and the serving system of Step-3~\cite{step-3} have also explored disaggregating attention and MoE layers into separate tiers. However, their approaches rely on CPU-controlled communication libraries based on technologies like GDRCopy. This reliance on the CPU for managing communication introduces additional overhead and prevents the end-to-end capture of CUDA graph, and further limiting potential optimizations, despite their disaggregated architecture. 

\paragraph{Communciation Libraries}
Efficient and stable communication libraries are imperative to support these parallelism strategies. NCCL~\cite{nccl} is widely used in both model training and inference for its ease of integration, while NVSHEMEM~\cite{nvshmem} provides scalable and efficient communication for NVIDIA GPUs. Moreover, systems such as Centauri~\cite{centauri} and Concerto~\cite{concerto} have demonstrated that overlapping techniques can reduce distributed communication overhead. In our work, we introduce a novel asymmetric asynchronous peer-to-peer communication library that supports our decoupled MoE servers by offering CPU-free communication with IBGDA~\cite{IBGDA}.

\section*{Conclusion}
We presented \ourmethod{}, a serving system designed for large-scale Mixture-of-Experts (MoE) models. By disaggregating MoE layers into independent services, \ourmethod{} achieves fine-grained elasticity, robust fault tolerance, and high efficiency. At its core, \ourmethod{} introduces a CPU-free, asynchronous P2P communication library that provides additional acceleration compared with existing solutions. Our evaluation demonstrates that \ourmethod{} delivers competitive throughput and latency compared to state-of-the-art systems while maintaining resilience under failures and strong scalability. This service-oriented paradigm establishes a flexible and efficient foundation for deploying large-scale MoE models in production.

\bibliography{MAIN}


\begin{thebibliography}{36}


\ifx \showCODEN    \undefined \def \showCODEN     #1{\unskip}     \fi
\ifx \showISBNx    \undefined \def \showISBNx     #1{\unskip}     \fi
\ifx \showISBNxiii \undefined \def \showISBNxiii  #1{\unskip}     \fi
\ifx \showISSN     \undefined \def \showISSN      #1{\unskip}     \fi
\ifx \showLCCN     \undefined \def \showLCCN      #1{\unskip}     \fi
\ifx \shownote     \undefined \def \shownote      #1{#1}          \fi
\ifx \showarticletitle \undefined \def \showarticletitle #1{#1}   \fi
\ifx \showURL      \undefined \def \showURL       {\relax}        \fi
\providecommand\bibfield[2]{#2}
\providecommand\bibinfo[2]{#2}
\providecommand\natexlab[1]{#1}
\providecommand\showeprint[2][]{arXiv:#2}

\bibitem[Chen et~al\mbox{.}(2024)]%
        {centauri}
\bibfield{author}{\bibinfo{person}{Chang Chen}, \bibinfo{person}{Xiuhong Li}, \bibinfo{person}{Qianchao Zhu}, \bibinfo{person}{Jiangfei Duan}, \bibinfo{person}{Peng Sun}, \bibinfo{person}{Xingcheng Zhang}, {and} \bibinfo{person}{Chao Yang}.} \bibinfo{year}{2024}\natexlab{}.
\newblock \showarticletitle{Centauri: Enabling Efficient Scheduling for Communication-Computation Overlap in Large Model Training via Communication Partitioning}. In \bibinfo{booktitle}{\emph{Proceedings of the 29th ACM International Conference on Architectural Support for Programming Languages and Operating Systems, Volume 3}} (La Jolla, CA, USA) \emph{(\bibinfo{series}{ASPLOS '24})}. \bibinfo{publisher}{Association for Computing Machinery}, \bibinfo{address}{New York, NY, USA}, \bibinfo{pages}{178–191}.
\newblock
\showISBNx{9798400703867}
\href{https://doi.org/10.1145/3620666.3651379}{doi:\nolinkurl{10.1145/3620666.3651379}}


\bibitem[Cheng et~al\mbox{.}(2025)]%
        {concerto}
\bibfield{author}{\bibinfo{person}{Shenggan Cheng}, \bibinfo{person}{Shengjie Lin}, \bibinfo{person}{Lansong Diao}, \bibinfo{person}{Hao Wu}, \bibinfo{person}{Siyu Wang}, \bibinfo{person}{Chang Si}, \bibinfo{person}{Ziming Liu}, \bibinfo{person}{Xuanlei Zhao}, \bibinfo{person}{Jiangsu Du}, \bibinfo{person}{Wei Lin}, {and} \bibinfo{person}{Yang You}.} \bibinfo{year}{2025}\natexlab{}.
\newblock \showarticletitle{Concerto: Automatic Communication Optimization and Scheduling for Large-Scale Deep Learning}. In \bibinfo{booktitle}{\emph{Proceedings of the 30th ACM International Conference on Architectural Support for Programming Languages and Operating Systems, Volume 1}} (Rotterdam, Netherlands) \emph{(\bibinfo{series}{ASPLOS '25})}. \bibinfo{publisher}{Association for Computing Machinery}, \bibinfo{address}{New York, NY, USA}, \bibinfo{pages}{198–213}.
\newblock
\showISBNx{9798400706981}
\href{https://doi.org/10.1145/3669940.3707223}{doi:\nolinkurl{10.1145/3669940.3707223}}


\bibitem[deepseek ai(2025)]%
        {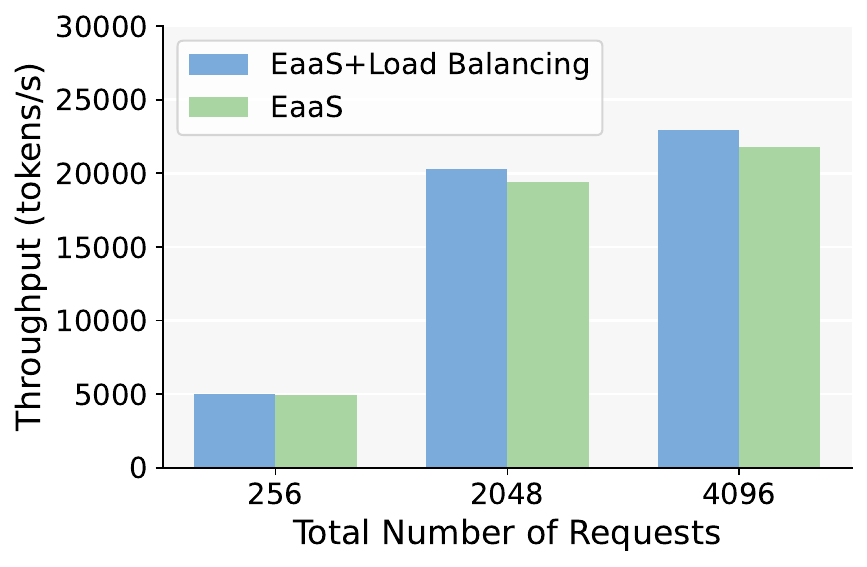}
\bibfield{author}{\bibinfo{person}{deepseek ai}.} \bibinfo{year}{2025}\natexlab{}.
\newblock \bibinfo{title}{Expert Parallelism Load Balancer (EPLB)}.
\newblock
\urldef\tempurl%
\url{https://github.com/deepseek-ai/EPLB}
\showURL{%
\tempurl}


\bibitem[Fedus et~al\mbox{.}(2022)]%
        {switch-transformers}
\bibfield{author}{\bibinfo{person}{William Fedus}, \bibinfo{person}{Barret Zoph}, {and} \bibinfo{person}{Noam Shazeer}.} \bibinfo{year}{2022}\natexlab{}.
\newblock \showarticletitle{Switch transformers: Scaling to trillion parameter models with simple and efficient sparsity}.
\newblock \bibinfo{journal}{\emph{Journal of Machine Learning Research}} \bibinfo{volume}{23}, \bibinfo{number}{120} (\bibinfo{year}{2022}), \bibinfo{pages}{1--39}.
\newblock


\bibitem[Guo et~al\mbox{.}(2025)]%
        {deepseek-r1}
\bibfield{author}{\bibinfo{person}{Daya Guo}, \bibinfo{person}{Dejian Yang}, \bibinfo{person}{Haowei Zhang}, \bibinfo{person}{Junxiao Song}, \bibinfo{person}{Ruoyu Zhang}, \bibinfo{person}{Runxin Xu}, \bibinfo{person}{Qihao Zhu}, \bibinfo{person}{Shirong Ma}, \bibinfo{person}{Peiyi Wang}, \bibinfo{person}{Xiao Bi}, {et~al\mbox{.}}} \bibinfo{year}{2025}\natexlab{}.
\newblock \showarticletitle{Deepseek-r1: Incentivizing reasoning capability in llms via reinforcement learning}.
\newblock \bibinfo{journal}{\emph{arXiv preprint arXiv:2501.12948}} (\bibinfo{year}{2025}).
\newblock


\bibitem[Huang et~al\mbox{.}(2018)]%
        {gpipe}
\bibfield{author}{\bibinfo{person}{Yanping Huang}, \bibinfo{person}{Youlong Cheng}, \bibinfo{person}{Ankur Bapna}, \bibinfo{person}{Orhan Firat}, \bibinfo{person}{Mia~Xu Chen}, \bibinfo{person}{Dehao Chen}, \bibinfo{person}{HyoukJoong Lee}, \bibinfo{person}{Jiquan Ngiam}, \bibinfo{person}{Quoc~V. Le}, \bibinfo{person}{Yonghui Wu}, {and} \bibinfo{person}{Zhifeng Chen}.} \bibinfo{year}{2018}\natexlab{}.
\newblock \bibinfo{title}{GPipe: Efficient Training of Giant Neural Networks using Pipeline Parallelism}.
\newblock
\href{https://doi.org/10.48550/ARXIV.1811.06965}{doi:\nolinkurl{10.48550/ARXIV.1811.06965}}


\bibitem[Hunt et~al\mbox{.}(2010)]%
        {ZooKeeper}
\bibfield{author}{\bibinfo{person}{Patrick Hunt}, \bibinfo{person}{Mahadev Konar}, \bibinfo{person}{Flavio~P. Junqueira}, {and} \bibinfo{person}{Benjamin Reed}.} \bibinfo{year}{2010}\natexlab{}.
\newblock \showarticletitle{ZooKeeper: wait-free coordination for internet-scale systems}. In \bibinfo{booktitle}{\emph{Proceedings of the 2010 USENIX Conference on USENIX Annual Technical Conference}} (Boston, MA) \emph{(\bibinfo{series}{USENIXATC'10})}. \bibinfo{publisher}{USENIX Association}, \bibinfo{address}{USA}, \bibinfo{pages}{11}.
\newblock


\bibitem[Jain et~al\mbox{.}(2020)]%
        {gems}
\bibfield{author}{\bibinfo{person}{Arpan Jain}, \bibinfo{person}{Ammar~Ahmad Awan}, \bibinfo{person}{Asmaa~M. Aljuhani}, \bibinfo{person}{Jahanzeb~Maqbool Hashmi}, \bibinfo{person}{Quentin~G. Anthony}, \bibinfo{person}{Hari Subramoni}, \bibinfo{person}{Dhableswar~K. Panda}, \bibinfo{person}{Raghu Machiraju}, {and} \bibinfo{person}{Anil Parwani}.} \bibinfo{year}{2020}\natexlab{}.
\newblock \showarticletitle{GEMS: GPU-Enabled Memory-Aware Model-Parallelism System for Distributed DNN Training}. In \bibinfo{booktitle}{\emph{SC20: International Conference for High Performance Computing, Networking, Storage and Analysis}}. \bibinfo{pages}{1--15}.
\newblock
\href{https://doi.org/10.1109/SC41405.2020.00049}{doi:\nolinkurl{10.1109/SC41405.2020.00049}}


\bibitem[Jiang et~al\mbox{.}(2024)]%
        {mixtral}
\bibfield{author}{\bibinfo{person}{Albert~Q Jiang}, \bibinfo{person}{Alexandre Sablayrolles}, \bibinfo{person}{Antoine Roux}, \bibinfo{person}{Arthur Mensch}, \bibinfo{person}{Blanche Savary}, \bibinfo{person}{Chris Bamford}, \bibinfo{person}{Devendra~Singh Chaplot}, \bibinfo{person}{Diego de~las Casas}, \bibinfo{person}{Emma~Bou Hanna}, \bibinfo{person}{Florian Bressand}, {et~al\mbox{.}}} \bibinfo{year}{2024}\natexlab{}.
\newblock \showarticletitle{Mixtral of experts}.
\newblock \bibinfo{journal}{\emph{arXiv preprint arXiv:2401.04088}} (\bibinfo{year}{2024}).
\newblock


\bibitem[Kwon et~al\mbox{.}(2023)]%
        {vllm}
\bibfield{author}{\bibinfo{person}{Woosuk Kwon}, \bibinfo{person}{Zhuohan Li}, \bibinfo{person}{Siyuan Zhuang}, \bibinfo{person}{Ying Sheng}, \bibinfo{person}{Lianmin Zheng}, \bibinfo{person}{Cody~Hao Yu}, \bibinfo{person}{Joseph Gonzalez}, \bibinfo{person}{Hao Zhang}, {and} \bibinfo{person}{Ion Stoica}.} \bibinfo{year}{2023}\natexlab{}.
\newblock \showarticletitle{Efficient Memory Management for Large Language Model Serving with PagedAttention}. In \bibinfo{booktitle}{\emph{Proceedings of the 29th Symposium on Operating Systems Principles}} (Koblenz, Germany) \emph{(\bibinfo{series}{SOSP '23})}. \bibinfo{publisher}{Association for Computing Machinery}, \bibinfo{address}{New York, NY, USA}, \bibinfo{pages}{611–626}.
\newblock
\showISBNx{9798400702297}
\href{https://doi.org/10.1145/3600006.3613165}{doi:\nolinkurl{10.1145/3600006.3613165}}


\bibitem[Li and Hoefler(2021)]%
        {li2021chimera}
\bibfield{author}{\bibinfo{person}{Shigang Li} {and} \bibinfo{person}{Torsten Hoefler}.} \bibinfo{year}{2021}\natexlab{}.
\newblock \showarticletitle{Chimera: efficiently training large-scale neural networks with bidirectional pipelines}. In \bibinfo{booktitle}{\emph{Proceedings of the International Conference for High Performance Computing, Networking, Storage and Analysis}}. \bibinfo{pages}{1--14}.
\newblock


\bibitem[Lin et~al\mbox{.}(2025)]%
        {Lin2025WeiPipe}
\bibfield{author}{\bibinfo{person}{Junfeng Lin}, \bibinfo{person}{Ziming Liu}, \bibinfo{person}{Yang You}, \bibinfo{person}{Jun Wang}, \bibinfo{person}{Weihao Zhang}, {and} \bibinfo{person}{Rong Zhao}.} \bibinfo{year}{2025}\natexlab{}.
\newblock \showarticletitle{WeiPipe: Weight Pipeline Parallelism for Communication-Effective Long-Context Large Model Training}. In \bibinfo{booktitle}{\emph{Proceedings of the 30th ACM SIGPLAN Annual Symposium on Principles and Practice of Parallel Programming}} (Las Vegas, NV, USA) \emph{(\bibinfo{series}{PPoPP '25})}. \bibinfo{publisher}{Association for Computing Machinery}, \bibinfo{address}{New York, NY, USA}, \bibinfo{pages}{225–238}.
\newblock
\showISBNx{9798400714436}
\href{https://doi.org/10.1145/3710848.3710869}{doi:\nolinkurl{10.1145/3710848.3710869}}


\bibitem[Liu et~al\mbox{.}(2024a)]%
        {deepseek-v2}
\bibfield{author}{\bibinfo{person}{Aixin Liu}, \bibinfo{person}{Bei Feng}, \bibinfo{person}{Bin Wang}, \bibinfo{person}{Bingxuan Wang}, \bibinfo{person}{Bo Liu}, \bibinfo{person}{Chenggang Zhao}, \bibinfo{person}{Chengqi Dengr}, \bibinfo{person}{Chong Ruan}, \bibinfo{person}{Damai Dai}, \bibinfo{person}{Daya Guo}, {et~al\mbox{.}}} \bibinfo{year}{2024}\natexlab{a}.
\newblock \showarticletitle{Deepseek-v2: A strong, economical, and efficient mixture-of-experts language model}.
\newblock \bibinfo{journal}{\emph{arXiv preprint arXiv:2405.04434}} (\bibinfo{year}{2024}).
\newblock


\bibitem[Liu et~al\mbox{.}(2024b)]%
        {deepseek-v3}
\bibfield{author}{\bibinfo{person}{Aixin Liu}, \bibinfo{person}{Bei Feng}, \bibinfo{person}{Bing Xue}, \bibinfo{person}{Bingxuan Wang}, \bibinfo{person}{Bochao Wu}, \bibinfo{person}{Chengda Lu}, \bibinfo{person}{Chenggang Zhao}, \bibinfo{person}{Chengqi Deng}, \bibinfo{person}{Chenyu Zhang}, \bibinfo{person}{Chong Ruan}, {et~al\mbox{.}}} \bibinfo{year}{2024}\natexlab{b}.
\newblock \showarticletitle{Deepseek-v3 technical report}.
\newblock \bibinfo{journal}{\emph{arXiv preprint arXiv:2412.19437}} (\bibinfo{year}{2024}).
\newblock


\bibitem[Liu et~al\mbox{.}(2023)]%
        {10.1145/3581784.3607073}
\bibfield{author}{\bibinfo{person}{Ziming Liu}, \bibinfo{person}{Shenggan Cheng}, \bibinfo{person}{Haotian Zhou}, {and} \bibinfo{person}{Yang You}.} \bibinfo{year}{2023}\natexlab{}.
\newblock \showarticletitle{Hanayo: Harnessing Wave-like Pipeline Parallelism for Enhanced Large Model Training Efficiency}. In \bibinfo{booktitle}{\emph{Proceedings of the International Conference for High Performance Computing, Networking, Storage and Analysis}} (Denver, CO, USA) \emph{(\bibinfo{series}{SC '23})}. \bibinfo{publisher}{Association for Computing Machinery}, \bibinfo{address}{New York, NY, USA}, Article \bibinfo{articleno}{56}, \bibinfo{numpages}{13}~pages.
\newblock
\showISBNx{9798400701092}
\href{https://doi.org/10.1145/3581784.3607073}{doi:\nolinkurl{10.1145/3581784.3607073}}


\bibitem[Meta(2025)]%
        {llama4}
\bibfield{author}{\bibinfo{person}{Meta}.} \bibinfo{year}{2025}\natexlab{}.
\newblock \showarticletitle{The Llama 4 herd: The beginning of a new era of natively multimodal AI innovation}.
\newblock \bibinfo{howpublished}{\url{https://ai.meta.com/blog/llama-4-multimodal-intelligence/}}.
\newblock  (\bibinfo{year}{2025}).
\newblock


\bibitem[{MLC team}(2023)]%
        {mlc-llm}
\bibfield{author}{\bibinfo{person}{{MLC team}}.} \bibinfo{year}{2023}\natexlab{}.
\newblock \bibinfo{booktitle}{\emph{{MLC-LLM}}}.
\newblock
\urldef\tempurl%
\url{https://github.com/mlc-ai/mlc-llm}
\showURL{%
\tempurl}


\bibitem[NVIDIA(2019)]%
        {cuda-graph}
\bibfield{author}{\bibinfo{person}{NVIDIA}.} \bibinfo{year}{2019}\natexlab{}.
\newblock \bibinfo{title}{Getting Started with CUDA Graphs}.
\newblock \bibinfo{howpublished}{\url{https://developer.nvidia.com/blog/cuda-graphs/}}.
\newblock


\bibitem[NVIDIA(2020)]%
        {nccl}
\bibfield{author}{\bibinfo{person}{NVIDIA}.} \bibinfo{year}{2020}\natexlab{}.
\newblock \bibinfo{title}{NVIDIA Collective Communications Library}.
\newblock
\urldef\tempurl%
\url{https://developer. nvidia.com/nccl}
\showURL{%
\tempurl}


\bibitem[NVIDIA(2022)]%
        {IBGDA}
\bibfield{author}{\bibinfo{person}{NVIDIA}.} \bibinfo{year}{2022}\natexlab{}.
\newblock \bibinfo{title}{Improving network performance of HPC systems using NVIDIA Magnum IO NVSHMEM and GPUDirect Async}.
\newblock \bibinfo{howpublished}{\url{https://developer.nvidia.com/blog/improving-network-performance-of-hpc-systems-using-nvidia-magnum-io-nvshmem-and-gpudirect-async}}.
\newblock


\bibitem[NVIDIA(2024)]%
        {trt-llm}
\bibfield{author}{\bibinfo{person}{NVIDIA}.} \bibinfo{year}{2024}\natexlab{}.
\newblock \bibinfo{booktitle}{\emph{TensorRT-LLM}}.
\newblock
\urldef\tempurl%
\url{https://github.com/NVIDIA/TensorRT-LLM}
\showURL{%
\tempurl}


\bibitem[NVIDIA(2025a)]%
        {GDRCopy}
\bibfield{author}{\bibinfo{person}{NVIDIA}.} \bibinfo{year}{2025}\natexlab{a}.
\newblock \bibinfo{title}{GDRCopy}.
\newblock \bibinfo{howpublished}{\url{https://github.com/NVIDIA/gdrcopy}}.
\newblock


\bibitem[NVIDIA(2025b)]%
        {nvshmem}
\bibfield{author}{\bibinfo{person}{NVIDIA}.} \bibinfo{year}{2025}\natexlab{b}.
\newblock \bibinfo{title}{NVSHMEM}.
\newblock \bibinfo{howpublished}{\url{https://developer.nvidia.com/nvshmem}}.
\newblock


\bibitem[OpenAI(2021)]%
        {openai-triton}
\bibfield{author}{\bibinfo{person}{OpenAI}.} \bibinfo{year}{2021}\natexlab{}.
\newblock \bibinfo{title}{Triton}.
\newblock \bibinfo{howpublished}{\url{https://github.com/triton-lang/triton}}.
\newblock


\bibitem[Qin et~al\mbox{.}(2025)]%
        {qin2025mooncakekvcachecentricdisaggregatedarchitecture}
\bibfield{author}{\bibinfo{person}{Ruoyu Qin}, \bibinfo{person}{Zheming Li}, \bibinfo{person}{Weiran He}, \bibinfo{person}{Mingxing Zhang}, \bibinfo{person}{Yongwei Wu}, \bibinfo{person}{Weimin Zheng}, {and} \bibinfo{person}{Xinran Xu}.} \bibinfo{year}{2025}\natexlab{}.
\newblock \bibinfo{title}{Mooncake: A KVCache-centric Disaggregated Architecture for LLM Serving}.
\newblock
\showeprint[arxiv]{2407.00079}~[cs.DC]
\urldef\tempurl%
\url{https://arxiv.org/abs/2407.00079}
\showURL{%
\tempurl}


\bibitem[Rajbhandari et~al\mbox{.}(2020)]%
        {rajbhandari2020zeromemoryoptimizationstraining}
\bibfield{author}{\bibinfo{person}{Samyam Rajbhandari}, \bibinfo{person}{Jeff Rasley}, \bibinfo{person}{Olatunji Ruwase}, {and} \bibinfo{person}{Yuxiong He}.} \bibinfo{year}{2020}\natexlab{}.
\newblock \bibinfo{title}{ZeRO: Memory Optimizations Toward Training Trillion Parameter Models}.
\newblock
\showeprint[arxiv]{1910.02054}~[cs.LG]
\urldef\tempurl%
\url{https://arxiv.org/abs/1910.02054}
\showURL{%
\tempurl}


\bibitem[Shoeybi et~al\mbox{.}(2020)]%
        {shoeybi2020megatronlmtrainingmultibillionparameter}
\bibfield{author}{\bibinfo{person}{Mohammad Shoeybi}, \bibinfo{person}{Mostofa Patwary}, \bibinfo{person}{Raul Puri}, \bibinfo{person}{Patrick LeGresley}, \bibinfo{person}{Jared Casper}, {and} \bibinfo{person}{Bryan Catanzaro}.} \bibinfo{year}{2020}\natexlab{}.
\newblock \bibinfo{title}{Megatron-LM: Training Multi-Billion Parameter Language Models Using Model Parallelism}.
\newblock
\showeprint[arxiv]{1909.08053}~[cs.CL]
\urldef\tempurl%
\url{https://arxiv.org/abs/1909.08053}
\showURL{%
\tempurl}


\bibitem[Team et~al\mbox{.}(2025)]%
        {kimi-k2}
\bibfield{author}{\bibinfo{person}{Kimi Team}, \bibinfo{person}{Yifan Bai}, \bibinfo{person}{Yiping Bao}, \bibinfo{person}{Guanduo Chen}, \bibinfo{person}{Jiahao Chen}, \bibinfo{person}{Ningxin Chen}, \bibinfo{person}{Ruijue Chen}, \bibinfo{person}{Yanru Chen}, \bibinfo{person}{Yuankun Chen}, \bibinfo{person}{Yutian Chen}, {et~al\mbox{.}}} \bibinfo{year}{2025}\natexlab{}.
\newblock \showarticletitle{Kimi k2: Open agentic intelligence}.
\newblock \bibinfo{journal}{\emph{arXiv preprint arXiv:2507.20534}} (\bibinfo{year}{2025}).
\newblock


\bibitem[Team(2023)]%
        {sharegpt}
\bibfield{author}{\bibinfo{person}{ShareGPT Team}.} \bibinfo{year}{2023}\natexlab{}.
\newblock \bibinfo{title}{ShareGPT}.
\newblock \bibinfo{howpublished}{\url{https://sharegpt.com/}}.
\newblock


\bibitem[Team(2025a)]%
        {dpattn}
\bibfield{author}{\bibinfo{person}{SGLang Team}.} \bibinfo{year}{2025}\natexlab{a}.
\newblock \bibinfo{title}{Data Parallelism Attention For DeepSeek Models}.
\newblock
\urldef\tempurl%
\url{https://lmsys.org/blog/2024-12-04-sglang-v0-4/#data-parallelism-attention-for-deepseek-models}
\showURL{%
\tempurl}


\bibitem[Team(2025b)]%
        {stepfun2025step3largeaffordablemodelsystem}
\bibfield{author}{\bibinfo{person}{StepFun Team}.} \bibinfo{year}{2025}\natexlab{b}.
\newblock \bibinfo{title}{Step-3 is Large yet Affordable: Model-system Co-design for Cost-effective Decoding}.
\newblock
\showeprint[arxiv]{2507.19427}~[cs.LG]
\urldef\tempurl%
\url{https://arxiv.org/abs/2507.19427}
\showURL{%
\tempurl}


\bibitem[Wang et~al\mbox{.}(2025)]%
        {step-3}
\bibfield{author}{\bibinfo{person}{Bin Wang}, \bibinfo{person}{Bojun Wang}, \bibinfo{person}{Changyi Wan}, \bibinfo{person}{Guanzhe Huang}, \bibinfo{person}{Hanpeng Hu}, \bibinfo{person}{Haonan Jia}, \bibinfo{person}{Hao Nie}, \bibinfo{person}{Mingliang Li}, \bibinfo{person}{Nuo Chen}, \bibinfo{person}{Siyu Chen}, {et~al\mbox{.}}} \bibinfo{year}{2025}\natexlab{}.
\newblock \showarticletitle{Step-3 is Large yet Affordable: Model-system Co-design for Cost-effective Decoding}.
\newblock \bibinfo{journal}{\emph{arXiv preprint arXiv:2507.19427}} (\bibinfo{year}{2025}).
\newblock


\bibitem[Yang et~al\mbox{.}(2025)]%
        {qwen3}
\bibfield{author}{\bibinfo{person}{An Yang}, \bibinfo{person}{Anfeng Li}, \bibinfo{person}{Baosong Yang}, \bibinfo{person}{Beichen Zhang}, \bibinfo{person}{Binyuan Hui}, \bibinfo{person}{Bo Zheng}, \bibinfo{person}{Bowen Yu}, \bibinfo{person}{Chang Gao}, \bibinfo{person}{Chengen Huang}, \bibinfo{person}{Chenxu Lv}, {et~al\mbox{.}}} \bibinfo{year}{2025}\natexlab{}.
\newblock \showarticletitle{Qwen3 technical report}.
\newblock \bibinfo{journal}{\emph{arXiv preprint arXiv:2505.09388}} (\bibinfo{year}{2025}).
\newblock


\bibitem[Zhao et~al\mbox{.}(2023)]%
        {zhao2023pytorchfsdpexperiencesscaling}
\bibfield{author}{\bibinfo{person}{Yanli Zhao}, \bibinfo{person}{Andrew Gu}, \bibinfo{person}{Rohan Varma}, \bibinfo{person}{Liang Luo}, \bibinfo{person}{Chien-Chin Huang}, \bibinfo{person}{Min Xu}, \bibinfo{person}{Less Wright}, \bibinfo{person}{Hamid Shojanazeri}, \bibinfo{person}{Myle Ott}, \bibinfo{person}{Sam Shleifer}, \bibinfo{person}{Alban Desmaison}, \bibinfo{person}{Can Balioglu}, \bibinfo{person}{Pritam Damania}, \bibinfo{person}{Bernard Nguyen}, \bibinfo{person}{Geeta Chauhan}, \bibinfo{person}{Yuchen Hao}, \bibinfo{person}{Ajit Mathews}, {and} \bibinfo{person}{Shen Li}.} \bibinfo{year}{2023}\natexlab{}.
\newblock \bibinfo{title}{PyTorch FSDP: Experiences on Scaling Fully Sharded Data Parallel}.
\newblock
\showeprint[arxiv]{2304.11277}~[cs.DC]
\urldef\tempurl%
\url{https://arxiv.org/abs/2304.11277}
\showURL{%
\tempurl}


\bibitem[Zheng et~al\mbox{.}(2024)]%
        {sglang}
\bibfield{author}{\bibinfo{person}{Lianmin Zheng}, \bibinfo{person}{Liangsheng Yin}, \bibinfo{person}{Zhiqiang Xie}, \bibinfo{person}{Chuyue Sun}, \bibinfo{person}{Jeff Huang}, \bibinfo{person}{Cody~Hao Yu}, \bibinfo{person}{Shiyi Cao}, \bibinfo{person}{Christos Kozyrakis}, \bibinfo{person}{Ion Stoica}, \bibinfo{person}{Joseph~E. Gonzalez}, \bibinfo{person}{Clark Barrett}, {and} \bibinfo{person}{Ying Sheng}.} \bibinfo{year}{2024}\natexlab{}.
\newblock \showarticletitle{{SGL}ang: Efficient Execution of Structured Language Model Programs}. In \bibinfo{booktitle}{\emph{The Thirty-eighth Annual Conference on Neural Information Processing Systems}}.
\newblock
\urldef\tempurl%
\url{https://openreview.net/forum?id=VqkAKQibpq}
\showURL{%
\tempurl}


\bibitem[Zhu et~al\mbox{.}(2025)]%
        {megascale-infer}
\bibfield{author}{\bibinfo{person}{Ruidong Zhu}, \bibinfo{person}{Ziheng Jiang}, \bibinfo{person}{Chao Jin}, \bibinfo{person}{Peng Wu}, \bibinfo{person}{Cesar~A. Stuardo}, \bibinfo{person}{Dongyang Wang}, \bibinfo{person}{Xinlei Zhang}, \bibinfo{person}{Huaping Zhou}, \bibinfo{person}{Haoran Wei}, \bibinfo{person}{Yang Cheng}, \bibinfo{person}{Jianzhe Xiao}, \bibinfo{person}{Xinyi Zhang}, \bibinfo{person}{Lingjun Liu}, \bibinfo{person}{Haibin Lin}, \bibinfo{person}{Li-Wen Chang}, \bibinfo{person}{Jianxi Ye}, \bibinfo{person}{Xiao Yu}, \bibinfo{person}{Xuanzhe Liu}, \bibinfo{person}{Xin Jin}, {and} \bibinfo{person}{Xin Liu}.} \bibinfo{year}{2025}\natexlab{}.
\newblock \bibinfo{title}{MegaScale-Infer: Serving Mixture-of-Experts at Scale with Disaggregated Expert Parallelism}.
\newblock
\showeprint[arxiv]{2504.02263}~[cs.DC]
\urldef\tempurl%
\url{https://arxiv.org/abs/2504.02263}
\showURL{%
\tempurl}


\end{thebibliography}

\end{document}